\documentclass{LMCS}

\usepackage{amsmath}
\usepackage{amssymb}
\usepackage{enumerate,hyperref}
\newcommand{\nat}{\mathit{Nat}}

\newcommand{\tuple}[1]{\langle#1\rangle}
\newcommand{\aut}{\mathcal{A}}
\newcommand{\bP}{\mathbf{P}}
\newcommand{\bQ}{\mathbf{Q}}
\newcommand{\bR}{\mathbf{R}}
\newcommand{\bX}{\mathbf{X}}
\newcommand{\bY}{\mathbf{Y}}
\newcommand{\bW}{\mathbf{W}}
\newcommand{\bU}{\mathbf{U}}

\newcommand{\bS}{\mathbf{S}}

\newcommand{\MLO}{\mathit{MLO}}
\newcommand{\col}{\mathit{col}}
\newcommand{\modP}{\nat,<,\bP}

\def\ra{{\rightarrow}}

\newcommand{\play}{\mathit{Play}}

\theoremstyle{plain}
\newtheorem{theorem}{Theorem}[section]
\newtheorem{theor}{Theorem}[section]
\newtheorem{lemma}[theorem]{Lemma}
\newtheorem{corol}[theorem]{Corollary}
\theoremstyle{definition}
\newtheorem{definition}[theorem]{Definition}
\newtheorem{observation}[theorem]{Observation}

\def\rar{\rightarrow}

\def\vphi{\varphi}
\newcommand{\mathdef}[1]{\relax\ifmmode #1\else $#1$\fi}

\newcommand{\autstates}{{\mathcal Q}}

\newcommand{\fair}{{\mathcal F}}

\newcommand{\auto}{{\mathcal A}}
\newcommand{\tra}{\delta_{\auto}}
\newcommand{\INF}{{\mathtt{Inf}}}
\newcommand{\win}{\mathit{WinSt}^I}
\newcommand{\loose}{\mathit{WinSt}^{II}}
\newcommand{\Wing}{\mathit{win}\mathit{St}^I_{G_\aut}}
\newcommand{\Loseg}{\mathit{win}\mathit{St}^{II}_{G_\aut}}
\newcommand{\WIN}{\mathit{WIN}^{II}}
\newcommand{\qinit}{q_\mathit{init}}
\newcommand{\dinit}{\delta_\mathit{init}}
\newcommand{\sigmout}{\Delta}
\newcommand{\out}{\mathit{out}}

\def\chM{M=\langle \nat,<, \bP_1, \dots, \bP_n\rangle }
\def\nek{\ldots}

\def\Models{\models}

\newcommand{\retrospective}{C-}
\newcommand{\sco}{SC-}
\newcommand{\fAct}{\mathbf{Fac}}

\def\eqalign#1{\null\,\vcenter{\openup\jot\mathsurround=0 pt
  \ialign{\strut\hfil$\displaystyle{##}$&$\displaystyle{{}##}$\hfil
      \crcr#1\crcr}}\,}
\def\:{\mathbin:}

\def\vp{\varphi}
\def\mG{\mathcal{G}}

\def\s{\subseteq}
\newcommand{\om}{\omega}
\newcommand{\opI}{\mathit{op}^I_{\aut}}
\newcommand{\opS}{\mathit{op}^{II}_{\aut}}

\def\doi{3 (4:9) 2007}
\lmcsheading%
{\doi}
{1--24}
{}
{}
{Jan.~\phantom{2}, 2007}
{Nov.~14, 2007}
{}

\begin{document}
\title{The Church Synthesis Problem with Parameters}

\author[A.~ Rabinovich]{Alexander Rabinovich}   
\address{Sackler Faculty of Exact  Sciences, Tel Aviv University, Israel 69978.} 

\email{rabinoa@post.tau.ac.il}

\keywords{Synthesis Problem, Decidability, Monadic Logic}
\subjclass{F.4.1;F.4.3}

\begin{abstract}
For a two-variable formula $\psi(X,Y)$ of Monadic Logic of Order
($\MLO$) the Church Synthesis Problem concerns the existence and
construction of an operator $Y=F(X)$ such that $\psi(X,F(X))$ is
universally valid over $\nat$.

B\"{u}chi and Landweber proved that the Church synthesis problem is
decidable; moreover, they showed that if there is an operator $F$ that
solves the Church Synthesis Problem, then it can also be solved by an
operator defined by a finite state automaton or equivalently by an
$\MLO$ formula. We investigate a parameterized version of the Church
synthesis problem. In this version $\psi$ might contain as a parameter
a unary predicate $P$. We show that the Church synthesis problem for
$P$ is computable if and only if the monadic theory of
$\tuple{\nat,<,P}$ is decidable.  We prove that the
B\"{u}chi-Landweber theorem can be extended only to ultimately
periodic parameters. However, the $\MLO$-definability part of the
B\"{u}chi-Landweber theorem holds for the parameterized version of the
Church synthesis problem.
\end{abstract}

\maketitle
\section{Introduction}

Two fundamental results of classical automata theory are
decidability of the monadic second-order logic of order (MLO) over
 $\omega=(\nat,<)$ and computability of the Church synthesis problem.
These results have provided the underlying mathematical framework
for the development of formalisms for the description of
interactive systems and their desired properties,  the algorithmic
verification  and the automatic synthesis of correct
implementations from logical specifications, and  advanced
algorithmic techniques that are now embodied in industrial  tools
for verification and validation.

B\"{u}chi \cite{Bu} proved that the monadic theory of
$\omega=\tuple{\nat,<}$ is decidable. Even before the decidability
of the monadic theory of  $\omega$ has been proved, it was shown
that the expansions of $\omega$ by ``interesting" functions have
undecidable monadic theory. In particular, the monadic theory of
$\tuple{\nat,<,+}$ and the monadic theory of
$\tuple{\nat,<,\lambda x. 2\times x}$  are undecidable
\cite{Rob58,Tra61}. Therefore, most efforts to find decidable
expansions  of $\omega$ deal with   expansions  of $\omega$ by
monadic predicates.

Elgot and Rabin \cite{ER66} found many interesting predicates
$\bP$ for which   $\MLO$ over  $\tuple{\nat,<,\bP}$ is decidable.
Among these predicates are the set of factorial numbers
$\{n!\:n\in\nat\}$, the sets of $k$-th powers $\{n^k\:n\in
\nat\}$ and the sets $\{k^n\:n\in \nat\}$  (for $k\in \nat$ ).

The Elgot and Rabin method  has been generalized and sharpened
over the  years and their results were extended to a variety of
unary predicates (see e.g., \cite{Ch69,Th75,Sem,CT02}). In
\cite{R05,R07,RT06} we provided necessary and sufficient
conditions for the decidability of monadic (second-order) theory
of expansions of the linear order of the  naturals $\omega$ by
unary predicates.

Let Spec  be a specification language and Pr be an implementation
language. The synthesis problem for these languages is  stated as
follows: find whether for a given specification $S(I,O)\in$SPEC
there is a program  ${\mathcal P}$ which implements it, i.e.,
$\forall I(S(I,{\mathcal P}(I))$.

The specification language for the  Church Synthesis problem is
the Monadic second-order Logic of Order. An $\MLO$  formula
$\varphi(X,Y)$ specifies a binary relation on  subsets of $\nat$.
Note that every subset $P$ of $\nat$  is associated with its
characteristic $\omega$-string $u_P$ (where $u_P(i)=1$ if $i\in P$
and otherwise $u_P(i)=0$). Hence, $\varphi(X,Y)$  can be
considered as a specification of  a binary relation on
$\omega$-strings.

 As  implementations, Church considers functions from the
set $\{0,1\}^{\omega}$ of $\omega$-strings over $\{0,1\}$ to
$\{0,1\}^{\omega}$. Such functions are called operators.
 A machine that computes  an operator at every moment $t\in \nat$
 reads an input symbol $X(t)\in\{0,1\}$ and produces  an output symbol
 $Y(t)\in \{0,1\}$. Hence, the output  $Y(t)$ produced at   $t$
 depends only on inputs symbols $X(0),X(1) ,\dots,  X(t)$.
Such operators are called {\em causal} operators (\retrospective
operators); if the output  $Y(t)$ produced at   $t$
 depends only on inputs symbols $X(0),X(1) ,\dots,  X(t-1)$, the
 corresponding operator is called {\em strongly causal}
 (SC-operator). The sets of recursive causal  and strongly causal  operators are defined
 naturally;
 a \retrospective or a \sco operator   is  a {\em finite
 state} operator
  if it is computable by a finite state automaton (for precise
definitions, see Subsection \ref{sub:aut}).

 The
following problem is known as the Church Synthesis problem.
\[\framebox {
\begin{minipage}{0.95\hsize}
\noindent \hspace{2.0cm}{\bf Church Synthesis problem}\vskip4 pt
\noindent{\em Input:} an  $\MLO$ formula $\psi(X,Y)$.\\
\noindent{\em Task:} Check whether there is a \retrospective
operator $F$ such that\\  $\mbox{}$\hspace{1.0cm}$ \nat\models
\forall X \psi(X, F(X)) $
  and if so, construct this operator.
\end{minipage}
}\]
The Church Synthesis problem is much more difficult than
the decidability  problem for $\MLO$ over $\omega$. B\"{u}chi and
Landweber \cite{BL69} proved that the Church synthesis problem is
computable. Their main theorem  is stated as follows:
 \begin{theor}\label{th:bl} For every $\MLO$   formula $\psi(X,Y)$ either
 there is a finite state \retrospective operator $F$ such that $\nat\models \forall X \psi(X,
 F(X))$ or there is a finite state \sco operator
 $G$ such that $\nat\models \forall Y\neg\psi(G(Y),
 Y)$. Moreover, it is decidable which of these cases holds and a
 corresponding operator is computable from $\psi$.
\end{theor}
In this paper we  consider natural generalizations of the Church
Synthesis Problem over  expansions of $\omega$ by monadic
predicates, i.e., over the structures $\tuple{\nat,<,\bP}$.

For example, let $\fAct=\{n!\:n\in\nat\}$ be the set of factorial
numbers, and let $\varphi(X,Y,\fAct)$ be a formula which specifies
that $t\in Y$ iff $t\in \fAct$ and $ (t'\in X)\leftrightarrow (t'\in
\fAct)$ for all $t'\leq t$. It is easy to observe that there is no
finite state \retrospective operator $F$ such that $\forall
X\varphi(X,F(X),\fAct)$. However, there is a recursive
\retrospective operator $H$ such that $\forall
X\varphi(X,H(X),\fAct)$. It is also easy to construct a finite state
\retrospective operator $G(X, Z)$ such that $\forall
X\varphi(X,G(X,\fAct),\fAct)$. It was surprising for us to discover
that it is decidable whether for a formula $\psi(X,Y,\fAct)$ there
is a \retrospective operator $F$ such that $\forall
X\varphi(X,F(X),\fAct)$ and if such an operator exists, then it is
recursive and computable from $\psi$.

Here is the summary of our results. We investigate a parameterized
version of the Church synthesis problem. In this version $\psi$
might contain as a parameter a unary predicate $P$. Below five
synthesis problems with a parameter $\bP\subseteq \nat$ are
stated. We use capital italic letters for unary predicate names
and set variables and the corresponding bold letters for their
interpretation.
\[\framebox {
\begin{minipage}{0.95\hsize}
\noindent \hspace{2.0cm}{\bf Synthesis Problems for
$\bP\subseteq\nat$}\vskip4 pt
\noindent{\em Input:}  an  $\MLO$ formula $\psi(X,Y,P)$.\\
\noindent{\em Problem 1:} Check whether there is a \retrospective
operator $Y=F(X,P)$ such that \\  $\mbox{}$\hspace{1.0cm} $
\nat\models \forall X \psi(X, F(X,\bP),\bP) $
 and if there is such a recursive operator  \\  $\mbox{}$\hspace{1.0cm}-  construct it.\\
\noindent{\em Problem 2:} Check whether there is a recursive
\retrospective operator $Y=F(X,P)$ such \\
$\mbox{}$\hspace{1.0cm} that $  \nat\models \forall X \psi(X,
F(X,\bP),\bP) $
 and if so - construct this operator.\\
 \noindent{\em Problem 3:} Check whether there is a recursive \retrospective operator $Y=F(X)$ such \\  $\mbox{}$\hspace{1.0cm} that
$ \nat\models \forall X \psi(X, F(X),\bP) $
 and if so - construct this operator.\\
\end{minipage}
}\]

The next two problems are obtained from problems 2 an 3 when
``recursive" is replaced  by ``finite state".
\[\framebox {
\begin{minipage}{0.95\hsize}
\noindent \hspace{2.0cm}{\bf Synthesis Problems for
$\bP\subseteq\nat$}\vskip4 pt
\noindent{\em Input:}  an  $\MLO$ formula $\psi(X,Y,P)$.\\
\noindent{\em Problem 4:} Check whether there is a finite state
\retrospective operator $Y=F(X,P)$ such \\
$\mbox{}$\hspace{1.0cm} that $  \nat\models \forall X \psi(X,
F(X,\bP),\bP) $
 and if so - construct this operator.\\
 \noindent{\em Problem 5:} Check whether there is a finite state  \retrospective operator $Y=F(X)$ such \\  $\mbox{}$\hspace{1.0cm} that
$ \nat\models \forall X \psi(X, F(X),\bP) $
 and if so - construct this operator.
\end{minipage}
}\]

We show
\begin{theor}\label{th:intr1} Let $\bP$ be a subset of
$\nat$. The following conditions are equivalent:
\begin{enumerate}[\em(1)]
\item Problem 1 for $\bP$  is computable.
\item Problem 2 for  $\bP$  is computable.
\item Problem 3 for  $\bP$  is computable.
\item The monadic
theory of $\tuple{\nat,<,\bP}$ is decidable. \item \label{cond5}
 For every $\MLO$
formula $\psi(X,Y,P)$ either
 there is a recursive \retrospective operator $F$ such that $\nat\models \forall X \psi(X,
 F(X),\bP)$ or there is a recursive \sco operator
 $G$ such that $\nat\models \forall Y\neg\psi(G(Y),
 Y,\bP)$. Moreover, it is decidable which of these cases holds and
 the (description of the) corresponding operator is computable from $\psi$.
 \end{enumerate}
\end{theor}
The  more difficult part of this theorem is the implication
(4)$\Rightarrow$(5).
%

The trivial examples of predicates with decidable monadic theory
are ultimately periodic predicates. Recall that a predicate $\bP$
is ultimately periodic if there are $0<p,d\in\nat$ such that
$(n\in \bP \leftrightarrow n+p\in \bP)$ for all $n>d$. Ultimately
periodic predicates are $\MLO$-definable. Hence, for these
predicates computability of Problems 1-5 can be derived from
Theorem \ref{th:bl}.

We prove that the  B\"{u}chi-Landweber theorem can be extended
only to ultimately periodic parameters.
\begin{theor}
\label{th:only-triv}
 Let $\bP$ be a subset of $\nat$. The following
conditions are equivalent and imply computability of Problem 4:
\begin{enumerate}[\em(1)]
\item $\bP$ is ultimately periodic.

\item
For every $\MLO$   formula $\psi(X,Y,\mathit{P})$ either
 there is a finite state \retrospective operator $F$ such that $\nat\models \forall X \psi(X,
 F(X,\bP),\bP)$ or there is a finite state  \sco operator
 $G$  such that $\nat\models \forall Y\neg\psi(G(Y,\bP),
 Y,\bP)$.

\end{enumerate}
\end{theor}
In Problems 1-5 we restrict the computational complexity of the
C-operators (implementations) which meet specifications. Another
approach is to restrict their descriptive complexity. The finite
state operators are $\MLO$-definable in $\om=\tuple{\nat,<}$. An
operator $F$ is defined by a formula $\psi(X,Y,P)$ in an expansion
$M=\tuple{\nat,<,\bP}$ of $\omega$, if for all $\om$-strings $\bX$
and $\bY$:
$$\bY=F(\bX) \mbox{ iff } \om\models\psi(\bX,\bY,\bP)$$
An operator $F$ is $\MLO$-definable  in  $M=\tuple{\nat,<,\bP}$,
 if it is defined by an $MLO$ formula in $M$

 Our main
theorem which is stated in the next section implies
 \begin{theor}\label{th:int4}  For every $\MLO$   formula $\psi(X,Y,P)$  and every expansion
$M\!=\!\tuple{\nat,<,\bP}$ of $\omega$  either
 there is an $\MLO$-definable (in $M$) \retrospective operator $F$ such that $M\models \forall X \psi(X,
 F(X),P)$ or there is an $\MLO$-definable (in $M$)  \sco operator
 $G$ such that $M\models \forall Y\neg\psi(G(Y),
 Y,P)$. Moreover,  formulas which define these operators are
 computable from $\psi$
\end{theor}

%
%
%
%
The paper is organized as follows. In the next section games and
their connections to the Church synthesis problem are discussed,
the  B\"{u}chi and Landweber  theorem is rephrased   in the game
theoretical language, and our main definability result - Theorem
\ref{th:main-new} - which implies Theorem \ref{th:int4} is stated.

In Section \ref{sec:th2},   Theorem
 \ref{th:intr1} is derived as a consequence of Theorem \ref{th:main-new}.
In Section \ref{sec:prel},  standard definitions and facts about
automata and logic are recalled. In Section \ref{sec:5}, finite
state synthesis problems with parameters
 are considered and Theorem \ref{th:only-triv} is proved.

In Section \ref{sec:games}, parity games on graphs  and their
connection to the synthesis problems are discussed and
definability results needed in the proof of  Theorem
\ref{th:main-new} are proved. The proof of Theorem
\ref{th:main-new}  is given in Section \ref{sec:th-main-new}.

Finally,  in Section \ref{sec:conc}, some open problems are stated
 and further results  are
 discussed.

\section{Game Version of  the Church Problem and Main Definability Result}

Let  $\bW$ be a set of pairs of $\omega$ strings over $\{0,1\}$. A
game $G(\bW)$ is defined as follows.
\begin{enumerate}[(1)]
\item The game is played by two players, called Player I (or Mr. $X$)  and Player II (or Mr. $Y$).
\item A \emph{play} of the game has $\omega$  rounds.
\item At round $n$:
first, Player I  chooses $x(n)\in \{0,1\}$; then, Player II
chooses $y(n)\in  \{0,1\}$.
\item By the end of the play the $\omega$ strings $\bY=y(0)y(1) \dots $ and $\bX=x(0)x(1) \dots $ over $\{0,1\}$
have been constructed.
\end{enumerate}

\begin{enumerate}[{}]
\item{\bf Winning conditions:} Player II  wins the play if the pair of $\om$-strings
$\tuple{\bX,\bY}$ is in $\bW$; otherwise, Player I wins the play.
\end{enumerate}

 What we
want to know is: Does either one of the players have a
\emph{winning strategy} in $G(\bW)$.  That is, can Player I choose
his moves so that in whatever way Player II responds, we have
$\tuple{\bX,\bY}\not\in \bW$? Or can Player II respond to Player
I's moves in a way that ensures the opposite?

Since at round $n$, Player I has access only to $\bY \cap [0,n)$
and Player II  has access only to $\bX\cap [0,n]$,
a strategy of Player I (respectively, of Player II) is a strongly
causal (respectively, causal) operator. So, a winning strategy for
Player II is a causal operator  $F:\{0,1\}^\om \ra \{0,1\}^\om$
 such that $\tuple{\bX,F(\bX)}$ is in $\bW$  for every $\bX \in \{0,1\}^\om
 $, and a winning strategy for
Player I is a strongly causal operator  $G:\{0,1\}^\om \ra
\{0,1\}^\om$
 such that $\tuple{G(\bY),\bY}$ is not in $\bW$  for every $\bY \in \{0,1\}^\om
 $.

 There is a natural topology on the set of $\omega$-strings (see
 e.g., \cite{PP04}). According to this topology,  subsets of
 $\om$-strings can be classified as open, closed, Borel and so
 forth.
 The following theorem of Martin is fundamental.
 \begin{theor}[Determinacy of Borel Games] \label{th:martin} For every Borel set
 $\bW$,
one of the players has a winning strategy in $G(\bW)$.
\end{theor}
It was McNaughton (see \cite{mcnaughton}) who first observed that
the Church problem  can be equivalently phrased in game-theoretic
language. Algorithmic questions deal with finitely described
objects.  Hence, McNaughton  considered games $G(\bW)$ only for
definable sets $\bW$.

Let  $M=\tuple{\modP} $ be a structure. A set
$\bW\s\{0,1\}^\om\times \{0,1\}^\om$ is \emph{defined by an $\MLO$
formula $\psi(X, Y,P) $  in $M$} if
$\bW=\{\tuple{\bX,\bY}\:M\models \psi(\bX,\bY,P)\}$. A set is
$\MLO$-\emph{definable} iff it is defined by an $\MLO$ formula.
Similarly,
 an operator  $F \:\{0,1\}^\om\ra \{0,1\}^\om$ is definable in $M$ if its graph
$\{\tuple{\bX,\bY}\:\bY=F(\bX)\}$ is definable; a strategy is
definable iff the corresponding causal operator is definable.

Let  $\chM$ be an expansion of $\om=\tuple{\nat,<}$ by  unary
predicates   and let $\psi(X,Y,P_1, \dots, P_n)$ be an $\MLO$
formula. The \emph{McNaughton game} $\mG_\psi^M$ is the game
$G(\bW)$, where $\bW$ is the set definable by $\psi$ in $M$.
Hence, the winning condition of $\mG_\psi^M$  can be stated as

\begin{enumerate}[{\ }]
\item{\bf Winning conditions for $\mG_\psi^M$:}   Player II  wins a play if $M\models
\psi(\bX,\bY,P_1,\dots, P_n)$;
 otherwise,
Player I wins the play.
\end{enumerate}

This leads to

\begin{enumerate}[{\ }]
\item{\bf Game version of the Church  problem:}\label{dfn:Church synthesis problem}
Let  $\chM$  be an expansion of $\om=\tuple{\nat,<}$ by  unary
predicates.
   Given a formula  $\psi(X,Y,P_1,\dots,P_n)$  decide
whether Player II  has a winning strategy in $\mG_\psi^M$.
\end{enumerate}

Theorem \ref{th:bl} states
the computability of the Church problem in the structure
$\om\!=\!\tuple{\nat,<}$ (no additional unary predicates). Even more
importantly, B\"{u}chi and Landweber show that in the case of
$\om$ we {can} restrict ourselves to
 \emph{definable strategies}, i.e., to   causal (or strongly causal)
operators computable by finite state automata or equivalently
$\MLO$-definable in $\om$. The B\"{u}chi-Landweber theorem can be
stated in the game theoretical language as follows:
\begin{theorem}[B\"{u}chi-Landweber, 1969]\label{th:bl-game} Let $\psi({X},{Y})$ be an $\MLO$  formula.,
Then:
\begin{enumerate}[{\ }]
\item{\bf Determinacy:} One of the players has a winning strategy in the game $\mG_\psi^\om$.
\item{\bf Decidability:} It is decidable \emph{which} of the players has a winning strategy.
\item{\bf Definable strategy:} The player who has a winning strategy also has an $\MLO$-\emph{definable} winning strategy.
\item{\bf Synthesis algorithm:} We can compute a formula $\vp({X},{Y})$ that defines (in $(\om,<)$)
a winning strategy for the winning player in $\mG_\psi^\om$.
\end{enumerate}
\end{theorem}
The determinacy part of the B\"{u}chi-Landweber theorem follows
from the determinacy of Borel games. More generally, it is well
known that for each $\MLO$ formula $\psi(X,Y,Z_1,\dots ,Z_n)$ and
$\bP_1,\dots,\bP_n\s\nat$ the set $\bW=\{\tuple{\bX,\bY}\:
\tuple{\nat,<}\models \psi(\bX,\bY,\bP_1,\dots,\bP_n)\}$ is a
Borel set (it is even inside the boolean closure of the second
level of the Borel hierarchy, see e.g., \cite{PP04}). Hence, by
the determinacy of Borel games, we obtain determinacy of the
McNaughton games. In other words, we have the following corollary:
\begin{corol}[Determinacy] \label{cor:martin}
Let $\chM$ be an expansion of $\omega$ by unary predicates. Then,
for every $\MLO$ formula $\psi(X,Y,P_1,\dots,P_n)$
\begin{enumerate}[\em(1)]
\item One of the players has a winning strategy in $\mG_\psi^M$.
\item Equivalently,
 either
 there is a \retrospective operator $F$ such that $M\models \forall X \psi(X,
 F(X))$, or there is a  \sco operator
 $G$ such that $M\models \forall Y\neg\psi(G(Y),
 Y)$.
 \end{enumerate}
\end{corol}
In order to simplify notations, from now on, we will state our
results only for the expansions of $\omega$ by one unary
predicate. The generalization to the expansions  by any number of
unary predicates is straightforward.

Now, we are ready to state our main result which generalizes the
definability and synthesis parts of  the B\"{u}chi-Landweber
theorem in a uniform way to the expansions of $\tuple{\nat,<}$ by
unary predicates.
\begin{theorem}[Main] \label{th:main-new} There is an algorithm that given a formula
$\vp(X,Y,P)$ constructs a sentence  $\mathit{WIN^{II}_\vp}(P)$ and
formulas $\mathit{St^I_\vp}(X,Y,P)$, $\mathit{St^{II}_\vp}(X,Y,P)$
such that for every structure $M= \tuple{\modP}$ Player II wins
the games $\mG_\vp^M$ iff $M\models \mathit{WIN^{II}_\vp}$.
Moreover, if Player II wins $\mG_\vp^M$,  then
$\mathit{St^{II}_\vp}(X,Y,P)$ defines his winning strategy;
otherwise, $\mathit{St^{I}_\vp}$ defines a winning strategy of
Player I.
\end{theorem}
 Theorem \ref{th:int4} is reformulated  in the game
language as follows.


\begin{theorem}[Game version of Theorem \ref{th:int4}]\label{th:int4-game} For every $\MLO$
formula $\psi(X,Y,P)$  and every expansion $M=\tuple{\nat,<,\bP}$
of $\omega$:
\begin{enumerate}[{\ }]
\item{\bf Determinacy:} One of the players has a winning strategy in the game $\mG_\psi^M$.
\item{\bf Definable strategy:} The player who has a winning strategy also has an $\MLO$-definable (in $M$)  winning strategy.
\item{\bf Synthesis algorithm:} We can compute a formula that defines (in $M$)
a winning strategy for the winning player in $\mG_\psi^M$.
\end{enumerate}
\end{theorem}
The determinacy part  of Theorem \ref{th:int4-game} follows from
Corollary \ref{cor:martin}. Its definability and synthesis parts
are
 immediate consequences of Theorem
\ref{th:main-new}.
 In the next
section, Theorem \ref{th:intr1} is derived as another  consequence
of Theorem \ref{th:main-new} and of Corollary \ref{cor:martin}.

The proof of Theorem \ref{th:main-new}  will be given in Sect.
\ref{sec:th-main-new}. Section \ref{sec:prel} provides an
additional background on logic and automata, and Section
\ref{sec:games} prepares  definability results needed in the proof
of  Theorem \ref{th:main-new}.

\section{Proof of Theorem \ref{th:intr1}}\label{sec:th2}
In this section we prove Theorem \ref{th:intr1}. Its more
difficult part is the implication (4)$\Rightarrow$(5); the proof
of this implication is easily obtained from  Theorem
\ref{th:main-new}. The proof of the other equivalences of Theorem
\ref{th:intr1} uses only  determinacy and  some simple facts.

We start with the following simple Lemma:
\begin{lemma} \label{lem:triv-new}
Assume that  the monadic theory of $\chM$ is decidable
\begin{enumerate}[\em(1)]
\item  Every  set  $\bQ\s\nat $ definable in $M$ is recursive.
\item Every  $C$-operator  $F$  definable in $M$ is recursive.
\item There is an algorithm that computes a program for $\bQ$
(respectively, for $F$) from a formula which defines $\bQ$
(respectively, $F$).
\end{enumerate}

\end{lemma}
\proof  (1) Assume that $\bQ$ is defined in $M$ by a formula
$\psi(X)$. Each $j\in \nat$ is defined by an $\MLO$ formula
$\varphi_j(t)$. Hence, one can check whether $j\in \bQ$, by
testing whether the sentence $\exists X \exists t\varphi_j(t)
\wedge \psi(X)\wedge t\in X$ holds in $M$.

(2) Assume that  a $C$-operator  $F$ is  defined in $M$ by a
formula $\psi(X,Y)$. We have to show that there is an algorithm
which computes the $n$-th letter  of $F(\bX)$ from the first $n$
letters of $\bX$. Let $a=a_0\dots a_{n}\in \{0,1\}^*$. Define
$\zeta^{a}_i(t,X) $ as $\vp_i(t)\wedge t\in X$ if $a_i$ is 1, and
as $\vp_i(t)\wedge t\not\in X$ otherwise, where $\vp_i(t)$ is a
formula which defines a number $i$. Note that $M\models\exists
t_0\exists t_1\dots \exists
t_n\bigwedge_{i=0}^n\zeta^{a}_i(t_i,\bX)\big)$ iff the first $n+1$
letters of $\bX$ are $a_0\dots a_{n}$.

Now, if  $a_0\dots a_{n}\in \{0,1\}^*$  are  the first $n+1$
letters of $\bX$, then the $n+1$-th letter of $F(\bX)$ is 1 iff
$$\exists X\exists Y \exists t_0 \exists t_1\dots \exists
t_n\big(\psi(X,Y)\wedge\vp_n(t_n)\wedge t_n\in
Y\wedge\bigwedge_{i=0}^n\zeta^{a}_i(t_i,X)\big)$$ holds in $M$.

(3) follows from the proofs of (1) and (2).\qed

\begin{lemma}[Implication (4)$\Rightarrow$(5) of Theorem  \ref{th:intr1}] \label{th:rec}
Assume that  the monadic theory of $M=\tuple{\nat,<,\bP}$ is
decidable. Then for every $\MLO$   formula $\vp(X,Y,P)$ either
 there is a recursive \retrospective operator $F$ such that $\nat\models \forall X \vp(X,
 F(X),\bP)$ or there is a recursive \sco operator
 $G$ such that $\nat\models \forall Y\neg\vp(G(Y),
 Y,\bP)$. Moreover,  it is decidable which of these cases holds and
 the (description of the) corresponding operator is computable from $\vp$.
\end{lemma}
\proof
For a formula $\vp$, construct $\mathit{WIN^{II}_\vp}(P)$ and
formulas $\mathit{St^I_\vp}(X,Y,P)$,
$\mathit{St^{II}_\vp}(X,Y,P)$, as in Theorem \ref{th:main-new}.

By the  assumption that the monadic theory of $M$ is decidable, we
can check whether $\mathit{WIN^{II}_\vp}(P)$ holds in $M$.

If  $\mathit{WIN^{II}_\vp}(P)$ holds in $M$    then, by Theorem
\ref{th:main-new}, $\mathit{St^{II}_\vp}(X,Y,P)$ defines a winning
strategy $F$ for Player II. Hence, $F$ is C-operator and
$\nat\models \forall X \vp(X,
 F(X),\bP)$.
 By Lemma \ref{lem:triv-new}, $F$ is  recursive.

 If  $\mathit{WIN^{II}_\vp}(P)$ does not hold in $M$,   then  by Theorem
\ref{th:main-new}, $\mathit{St^{I}_\vp}(X,Y,P)$ defines a winning
strategy $G$ for Player I. Hence, $G$ is SC-operator and
$\nat\models \forall Y\neg\vp(G(Y),
 Y,\bP)$. Moreover, by Lemma \ref{lem:triv-new}, $G$ is  recursive.\qed

\begin{lemma} \label{lem:prob2-sem} If  one of the Problems 1-5 is computable for $\bP$, then the
monadic theory of $\tuple{\nat,<,\bP}$ is decidable.
\end{lemma}
\proof
Let $\beta(P)$ be a sentence in $\MLO$ and
 let $\psi_{\beta}(X,Y,P)$ be defined as $$\big(\beta\ra( Y=\{0\})\big)\wedge
 \big(\neg\beta \ra( X=\emptyset)\big).$$

Observe  that $\nat\models\beta(\bP)$ iff there is a \retrospective
operator $F$ such that:  $$ \nat\models\forall X
\psi_{\beta}(X,F(X,\bP),\bP)\mbox{ iff  } \nat\models\forall X
\psi_{\beta}(X,H(X,\bP),\bP)$$
 where $H$ is a constant  \retrospective operator
 defined as
$ H=\lambda \tuple{X,P}. 10^{\omega} $.

Hence, if one of the Problems 1-5 is computable for $\bP$, then we
can decide whether $\nat\models\beta(\bP)$.\qed

The proof of Lemma \ref{lem:prob2-sem} also implies that if the
following  Problem 1$'$ is decidable for $\bP$, then the monadic
theory of $\tuple{\nat,<,\bP}$ is decidable.
\[\framebox {
\begin{minipage}{0.95\hsize}
\noindent \hspace{2.0cm}{\bf Decision  Problem $1'$ for
$\bP\subseteq\nat$}\vskip4 pt
\noindent{\em Input:}  an  $\MLO$ formulas $\psi(X,Y,P)$.\\
\noindent{Question:} Check whether there is a \retrospective
operator $Y=F(X,P)$ such that \\  $\mbox{}$\hspace{1.4cm}  $
\nat\models \forall X \psi(X, F(X,\bP),\bP). $
\end{minipage}
}\]
Problem 1$'$ is actually Problem 1 without construction
part.

Finally, we have
 \begin{lemma} \label{lem:pr123} The implications
  (5)$\Rightarrow$(1),   (5)$\Rightarrow$(2) and (5)$\Rightarrow$(3) hold.
\end{lemma}
\proof

Let $\psi(X,Y,P)$ be a formula. By (5) either there  is a
recursive \retrospective operator $F$ such that $\nat\models
\forall X \psi(X,
 F(X),\bP)$ or there is a recursive \sco operator
 $G$ such that $\nat\models \forall Y\neg\psi(G(Y),
 Y,\bP)$. Moreover, it is decidable which of these cases holds and
 the corresponding operator is computable from $\psi$.

In the first case, the answer to Problems 1-3 is positive and $F$
is a corresponding operator.

In the second case, the answer to Problems 1-3 is negative.

 Indeed, for the sake of contradiction, assume that
 there is a \retrospective operator (even non-recursive) $F$ such that $\nat\models
\forall X \psi(X,
 F(X,\bP),\bP)$.
 Observe that $F$ is a \retrospective operator and $G$ is a \sco
 operator. Hence, $H=\lambda X. G(F(X,\bP))$ is a \sco operator.
 Every  \sco operator has a fixed point.
 Let
  $\mathbf{X}_0$  be a fixed point of $H$ and let  $\mathbf{Y}_0=F(\mathbf{X}_0,\bP)$.
  Then we have:  $\mathbf{X}_0=G(\mathbf{Y}_0)$. Therefore,
 we obtain
\[\eqalign{
  \nat\models\psi(\mathbf{X}_0,\mathbf{Y}_0,\bP)&\quad\hbox{because}\quad
  \nat\models\forall X \psi(X,F(X,\bP),\bP)\ ,\quad\hbox{and}\cr
  \nat\models\neg\psi(\mathbf{X}_0,\mathbf{Y}_0,\bP)&\quad\hbox{because}\quad
  \nat\models \forall Y\neg\psi(G(Y),Y,\bP)\ .
}
\]
  Contradiction.\qed

\section{Background on Logic and Automata} \label{sec:prel}
\subsection{ Notations and Terminology} \label{subsec:term}
 We use $k,~l,~m,~n,~i$ for
natural numbers;  $\nat$ for the set of natural numbers and
capital bold letters $\bP,~\bS,~\bR$ for subsets of $\nat$. We
identify subsets of a set $A$ and the corresponding unary
(monadic) predicates on $A$.

The set of  all (respectively, non-empty) finite strings over an
alphabet $\Sigma$ is denoted by  $\Sigma^*$ (respectively, by
$\Sigma^+$). The set of $\omega$-strings over $\Sigma$ is denoted
by $\Sigma^{\omega}$.

Let $a_0\dots a_k \dots $ and $b_0\dots b_k \dots $  be
$\omega$-strings. We say that these $\omega$-strings coincide on
an interval $[i,j]$ if $a_k=b_k$ for $i\leq k\leq j$.  A function
$F$ from $\Sigma_1^{\omega}$ to $\Sigma_2^{\omega}$ will be called
an operator of type $\Sigma_1\ra\Sigma_2$. An operator $F$
 is called \emph{causal} (respectively, \emph{strongly causal}) operator, if
$F(X)$ and $F(Y)$ coincide on an interval $[0,t]$, whenever $X$
and $Y$ coincide on $[0,t]$ (respectively, on $[0,t)$). We will
refer to causal (respectively, strongly causal) operators as
C-operators (respectively, SC-operators).

Let $\Sigma_1$  and $\Sigma_2$ be finite alphabet and let
$F\:\Sigma_1^{\omega}\ra\Sigma_2^{\omega}$ be a C-operator. Note
that there is a unique  function $h_F\:\Sigma_1^*\ra \Sigma_2$
such that $F(a_1 \dots a_n)=b_n$ if for some (equivalently for
all) $\omega$-string $y$:  $b_n$ is the $n$-th letter of $F(a_1
\dots a_ny)$. $F$ is said to be {\em recursive} if $h_F$ is
recursive.

Every \sco operator $F$ of type $\Sigma\ra\Sigma$ has a unique
fixed point, i.e., there is a unique $X\in \Sigma^{\omega}$ such
that $X=F(X)$.

Let $G:\Sigma^{\omega}\ra \sigmout^{\omega}$ be an operator. In
the case $\Sigma$ is the Cartesian product $\Sigma_1\times
\Sigma_2$ we will identify  $G$ with the corresponding operator
$F: \Sigma_1^{\omega} \times \Sigma_2^{\omega} \ra
\sigmout^{\omega}$. An operator $F: \Sigma_1^{\omega} \times
\Sigma_2^{\omega} \ra \sigmout^{\omega}$ is said to be \sco
operator (\retrospective operator) if $G$   is \sco operator
(respectively, \retrospective operator).

 There exists a one-one correspondence between the set of all
$\omega$-strings over  the alphabet $\{0,1\}^n$ and the set of all
$n$-tuples $\tuple{\bP_1,\dots,\bP_n}$ of unary predicates over
the set of natural numbers. With an $n$-tuple
$\tuple{\bP_1,\dots,\bP_n}$  of unary predicates over $\nat$, we
associate the $\omega$-string $a_0 a_1 \dots a_k \dots $ over
alphabet $\{0,1\}^n$ defined by $a_k =_{\mathit{def}} \langle
b^k_1, \dots b^k_n \rangle$ where $b^k_i$ is 1 if $\bP_i(k) $
holds and $b^k_i$ is 0 otherwise. Let  $Q=\{q_1,\dots , q_m\}$  be
 a finite  set of state.  There is a natural one-one
correspondence between  the subsets of $Q\times \nat$ and the set
of $m$-tuples of   unary predicates over $\nat$: with $U\subseteq
Q\times\nat$ we associate the $m$-tuple $\tuple{\bP_1, \dots
,\bP_m}$ defined as $i\in \bP_j$ iff $U(q_j,i)$ (for $i\in \nat$
and $j\leq m$).

Similarly, there is a one-one correspondence between the set of
all strings of length $m$  over  the alphabet $\{0,1\}^n$ and the
set of all $n$-tuples $\tuple{\bP_1,\dots,\bP_n}$ of unary
predicates over the set $\{0,\dots,m-1\}$.

A linearly ordered set will be called a chain. A chain with $n$
monadic predicates over its domain will be called an $n$-labelled
chain; whenever $n$ is clear from the context, $n$-labelled chains
will be called labelled  chains.

We  will sometimes  identify an $n$-labelled chain $\chM$  with
the $\omega$-string  over the alphabet $\{0,1\}^n$  which
corresponds to the $n$-tuple $\tuple{\bP_1,\dots,\bP_n}$; this
$\omega$-string will be called the {\em characteristic}
$\omega$-string (or $\omega$-word)  of $M$.
 Similarly, we will
identify finite $n$-labelled chains with corresponding strings
over $\{0,1\}^n$.
\subsection{Monadic Second-Order Logic and Monadic Logic of Order}
Let $\sigma$ be a  relational signature. Atomic formulas of the
monadic second-order logic over $\sigma$ are $R(t_1, . . . ,
t_n)$, $t_1 = t_2$, and $t_1\in X$ where $t_1,\dots ,t_n$ are
individual variables, $R\in \sigma$ is an n-are relational symbol,
and $X$ is a set variable. Formulas are obtained from atomic
formulas by conjunction, negation, and quantification $\exists t$
and $\exists X$ for $t$ an individual and $X$ a set variable. The
satisfaction relation $ M,\tau_1,\ldots \tau_k ;
\bS_1,\ldots,\bS_m \Models\varphi(t_1,\ldots, t_k ;X_1,\nek ,X_m)$
is defined as usual with the understanding that set variables
range over subsets of $M$.

We use standard abbreviations, e.g.,  we write $X \subseteq X'$
for $\forall t.~X(t)\ra X'(t)$;   we write $X =X'$ for $\forall
t.~X(t)\leftrightarrow X'(t)$; symbols ``$\exists^{\leq 1}$'' and
``$\exists!$'' stands for ``there is  at most one'' and ``there is
a unique''.

If a  signature $\sigma$  contains one binary predicate $<$ which
is interpreted as a linear order,  and all other predicates are
unary, the monadic second-order logic  for this signature is
called Monadic Logic of Order ($\MLO$). The formulas of $\MLO$ are
interpreted over labelled chains.

The {\em monadic theory} of a  labelled chain $M$ is the set of
all $\MLO$ sentences which hold in $M$.

 We will deal with the
   expansions of
$\omega$  by monadic predicates, i.e., with  the  structures of
the form $\chM$. We say that a chain $\chM$  is {\em recursive} if
all $\bP_i$ are recursive subsets of $\nat$.

An $\omega$-language $L$ is said to be \emph{defined} by an $\MLO$
formula $\psi(X_1,\dots, X_n)$ if the following condition holds:
an $\omega$ string is in $L$ iff the corresponding $n$-tuple of
unary predicates satisfies $\psi$.

\subsection{The First-Order Version of the Monadic Second-Order
Logic} \label{sec:fov} Sometimes it will be convenient for us to
consider the first-order version of the monadic second order
logic.

Let $\sigma$ be a  relational signature and $M$ be a structure for
$\sigma$.

Let $\bar{\sigma}=\sigma\cup\{Sing,\subseteq\}$ where $Sing$ is a
new unary relational symbol and $\subseteq $ a new binary
relational symbol.
 Let $\bar{M}$  be the structure for $\bar{\sigma} $ defined as follows:
 The domain  of  $\bar{M}$  is the set of all subsets of the domain of $M$.
 $Sing(A)$ holds in $\bar{M}$ if $A$ is one element subset; $A\subseteq B$ holds if $ A $ is a subset of $B$;
 for $k$-ary relational symbol $R\in \sigma$:
 $$R(A_1,\dots A_k) \mbox{ holds in } \bar{M}
 \mbox{ iff}$$
$$A_1=\{a_1\},\dots ,A_k=\{a_k\} \mbox { and } R(a_1,\dots a_k)\mbox{ holds in }M$$

The following lemma is well-known and is easily proved by the
structural induction.
\begin{lemma}[Equivalence of two Versions of Monadic Logic] \label{lem:fov-ofmso} The two versions of Monadic  logic are expressive equivalent, that is
\begin{enumerate}[\em(1)]
\item Let $\psi(X_1, \dots ,X_k)$ be a formula of the monadic
second-order logic for a signature $\sigma$. There is a
first-order formula $\vphi(X_1,\dots, X_k)$ in the signature
$\bar{\sigma}$ such that for every structure $M$ for the signature
$\sigma$ and for subsets $A_1,\dots, A_k$ of the domain of $M$
$$M,A_1,\dots, A_k\models \psi(X_1,\dots ,X_k)\mbox{ iff }
\bar{M},A_1,\dots, A_k\models \vphi(X_1,\dots ,X_k).$$ Moreover
there is an algorithm that computes $\vphi$ from $\psi$.
\item
Let $\vphi(X_1, \dots ,X_k)$ be a  first-order formula  in the
signature $\bar{\sigma}$. There is a formula in the monadic
second-order logic for the signature $\sigma$ such that for every
structure $M$ for the signature $\sigma$ and for subsets
$A_1,\dots, A_k$ of the domain of $M$
$$M,A_1,\dots, A_k\models \psi(X_1,\dots ,X_k)\mbox{ iff }
\bar{M},A_1,\dots, A_k\models \vphi(X_1,\dots ,X_k).$$ Moreover
there is an algorithm that computes $\psi$ from $\vphi$.\qed
\end{enumerate}
\end{lemma}

\subsection{Automata} \label{sub:aut}
A  \emph{deterministic transition system} $D$  is a tuple
$\tuple{\autstates,\Sigma,\delta, \qinit}$, consisting of a set
$\autstates$ of \emph{states}, an  \emph{alphabet} $\Sigma$, a
\emph{transition function} $\delta\:\autstates \times \Sigma\ra
\autstates$  and \emph{initial state} $\qinit\in \autstates$. The
transition function is extended as usual to a function from
$\autstates \times \Sigma^*$ to $\autstates $ which will be also
denoted by $\delta$. The function $\dinit\:\Sigma^*\ra
\autstates$ is defined as $\dinit(\pi)=\delta(\qinit,\pi)$. A
transition systems is finite if $\autstates$ and $\Sigma$ are
finite.

A  \emph{finite deterministic automaton} $\aut$  is a tuple
$\tuple{\autstates,\Sigma,\delta, \qinit,F}$, where
$\tuple{\autstates,\Sigma,\delta, \qinit}$ is a finite
deterministic transition system and $F$ is a subset of
$\autstates$. A string $\pi\in\Sigma^*$ is accepted by $\aut$ if
$\dinit(\pi)\in F$. The language accepted (or defined) by $\aut$
is the set of string accepted by $\aut$.

A \emph{Mealey automaton} is a tuple
$\tuple{\autstates,\Sigma,\delta, \qinit,\sigmout, \out }$, where
$\tuple{\autstates,\Sigma,\delta,  \qinit}$ is a deterministic
transition system, $\sigmout$  is an alphabet and
$\out\:\autstates\ra \sigmout$ is an output  function. With a
Mealey automaton ${\aut}=\tuple{\autstates,\Sigma,\delta,
\qinit,\sigmout, \out }$ we associate  a function
$h_{\aut}\:\Sigma^{*}\ra \sigmout$ and an  operator $F_{\aut}\:
\Sigma^{\omega}\ra \sigmout^{\omega}$ defined as follows:
$$h_{\aut}(a_0\dots a_{i-1})=\out(\dinit(a_0\dots a_{i-1}))$$
$$F_{\aut}(a_0\dots a_i\dots)= b_0\dots b_i\dots \mbox{ iff }
b_i=h_{\aut}(a_0\dots a_{i-1})$$ It is easy to see that an
operator is strongly causal (\sco operator) iff it is definable by
a Mealey automaton. We say that a \sco operator $F
\:\Sigma^{\omega}\ra \sigmout^{\omega}$  is  finite state iff it
is definable by a finite state Mealey automaton.

A finite Mealey automaton $\mathcal A=\tuple{\autstates,\Sigma,\rar,
\delta, \qinit, \Delta, {\col}}$, where the output alphabet $\Delta$
is a (finite) subset of $\nat$, is called a (deterministic)
\emph{parity automaton}; the output function $\col$ is usually refered
to as \emph{coloring function}.

With every $\omega$-string $a_0a_1 \cdots a_i\dots\in \Sigma^{\omega}$
we associate the $\omega$-sequence of successive states $\dinit(a_0)
\dinit(a_0a_1) \cdots \dinit(a_0 \cdots a_i)\cdots$ and the set $\INF$
of all $q\in \autstates$ that appear infinitely many times in this
sequence. An $\omega$-string is accepted by $\aut$ if the minimal
element of the set $\{\col(q) \:q\in \INF\}$ is even.  The
$\omega$-language {\em accepted} (or defined) by $\aut$ is the set of
all $\omega$-strings accepted by $\aut$.

 Sometimes the
alphabet $\Sigma$ of $\aut$ will be the Cartesian product $\Sigma_1
\times \Sigma_2\times \Sigma_3$ of other alphabets. In this case  we
say that  $\aut$  defines a relation $R_{\aut}\subseteq
\Sigma^{\omega}_1 \times \Sigma^{\omega}_2\times \Sigma^{\omega}_3
$; a triplet $\tuple{a,b,c}$ of $\omega$-strings is  in $R_{\aut}$
iff the $\omega$ string $(a_0,b_0,c_0)(a_1,b_1,c_1)\dots
(a_i,b_i,c_i)\dots $ is accepted by $\aut$.

Here is  the classical theorem due to B\"{u}chi, Elgot and
Trakhtenbrot.
\begin{theor}\label{th:log-aut}
\begin{enumerate}[\em(1)]
\item A language  is accepted by a finite deterministic  automaton iff it
is definable by an $\MLO$ formula.
\item An $\omega$-language is accepted by a deterministic parity automaton iff it
is definable by an $\MLO$ formula.
\item Moreover,
there is an algorithm which for every formula $\vphi(X_1,\dots,
X_m)$ computes an equivalent  deterministic
 automaton $\aut$ i.e., the language
definable by $\vphi$ is accepted by $\aut$.
 There is an algorithm which for every  deterministic
 automaton $\aut$ computes an equivalent $\MLO$ formula.
 Similarly, there are translation  algorithms  between formulas
 and deterministic parity automata.\qed
\end{enumerate}
\end{theor}
A \emph{Moore automaton} is a tuple
$\tuple{\autstates,\Sigma,\delta, \qinit,\sigmout, \out }$, where
$\tuple{\autstates,\Sigma,\delta, \qinit}$ is a deterministic
transition system, $\sigmout$ is an alphabet and
$\out\:\autstates \times \Sigma \ra \sigmout$ is an output
function.

With a Moore automaton ${\aut}=\tuple{\autstates,\Sigma,\delta,
\qinit,\sigmout, \out }$ we associate  a  function
$h_{\aut}\:\Sigma^{+}\ra \sigmout$ and  an operator $F_{\aut} \:
\Sigma^{\omega}\ra \sigmout^{\omega}$ defined as follows:
$$ h_{\aut}(a_0\dots , a_i)= \out(\dinit(a_0\dots a_{i-1}), a_i)$$
$$F_{\aut}(a_0\dots a_i\dots)= b_0\dots b_i\dots \mbox{ iff }
b_i=h_{\aut}(a_0\dots , a_i) $$
It is easy to see that an operator
is  causal (\retrospective operator) iff it is definable by a
Moore automaton.

We say that a \retrospective operator $F \:\Sigma^{\omega}\ra
\sigmout^{\omega}$  is  finite state iff
it is definable by a finite state 
Moore automaton.

\section{Finite State Synthesis Problems with Parameters}\label{sec:5}
 Recall that a predicate $\bP\subseteq \nat$
is ultimately periodic if there is $p,d\in\nat$ such that  $(n\in
\bP \leftrightarrow n+p\in \bP)$ for all $n>d$. Ultimately periodic
predicates are $\MLO$-definable. Therefore, for every ultimately
periodic predicate $\bP$ the  monadic theory of $\tuple{\nat,<,\bP}$
is decidable.

The next theorem implies Theorem \ref{th:only-triv}  and  shows
that Theorem \ref{th:bl} can be extended only to ultimately
periodic predicates.

\begin{theor}
\label{th:reduction}
 Let $\bP$ be a subset of $\nat$. The following
conditions are equivalent  and imply computability of Problem 4:
\begin{enumerate}[\em(1)]
\item $\bP$ is ultimately periodic.

\item
For every $\MLO$   formula $\psi(X,Y,\mathit{P})$ either
 there is a finite state \retrospective operator $F$ such that $\nat\models \forall X \psi(X,
 F(X,\bP),\bP)$ or there is a finite state  \retrospective operator
 $G$  such that $\nat\models \forall Y\neg\psi(G(Y,\bP),
 Y,\bP)$.
\item $\bP$ satisfies  the following selection
condition:
\begin{quote}
 For every formula $\alpha(X,\mathit{P})$ such that $\nat\models
\exists X\alpha(X,\bP)$ there is a finite state \retrospective
operator $H\: \{0,1\}^{\omega} \ra \{0,1\}^{\omega} $ such that
$\nat\models \alpha(H(\bP), \bP )$.
\end{quote}
\end{enumerate}
\end{theor}

\begin{proof}
The implication (1)$\Rightarrow(2)$ 
 follows from Theorem \ref{th:bl}  and the
fact that every ultimately periodic predicate is definable by an
$\MLO$ formula. The implication (2)$\Rightarrow$(3) is trivial.

The implication   (3)$\Rightarrow$(1) is derived as follows. Let
$\alpha(X,P)$ be $\forall t \big(X(t)\leftrightarrow P(t+1)\big)$.
Note  $\nat \models \exists X\alpha(X,\bP)$ for every
$\bP\subseteq \nat$. Therefore, if $\bP$ satisfies selection
condition, then there is \retrospective operator $H :
\{0,1\}^{\omega} \ra \{0,1\}^{\omega} $ such that $\nat\models
\alpha(H(\bP), \bP )$.

Assume  that a finite state Moore automaton   $\aut$ computes $H$
and has $n$ states. We are going to show that $\bP$ is ultimately
periodic with period at most $2n+1$. For $i\in \nat$ let $a_i$ be
one if $i\in\bP$  and $a_i$ be zero otherwise. Let $q_0q_1 \dots
q_{2n+1} \dots $ be the sequence states passed by $\aut$  on  the
input $a_0a_1 \dots a_{2n+1}\dots $. There are $i<j<2n$ such that
$a_i=a_j$ and $q_i=q_j$. Observe that $ q_{i+1}
=\delta_{\aut}(q_i,a_{i})= \delta_{\aut}(q_j,a_{j})=q_{j+1} $ and
$a_{i+1}=\out_{\aut}(q_{i},a_i)=\out_{\aut}(q_{j},a_j)=a_{j+1}$.
And by induction we get that $q_{i+m}=q_{j+m}$ and
$a_{i+m}=a_{j+m}$ for all $m\in\nat$. Therefore, $\bP$ is an
ultimately periodic with a period $j-i<2n$.
\end{proof}
Note that this theorem does not imply that Problem 4 is computable
only for ultimately periodic predicates.  The next theorem can be
established by the same  arguments.

\begin{theor}
\label{th:condprob5}

The following conditions are equivalent and imply computability of
Problem 5:
 \begin{enumerate}[\em(1)]
\item $\bP$ is ultimately periodic.
\item
For every $\MLO$   formula $\psi(X,Y,P)$ either
 there is a finite state \retrospective operator $F$ such that $\nat\models \forall X \psi(X,
 F(X),\bP)$ or there is a finite state \sco operator
 $G$ such that $\nat\models \forall Y\neg\psi(G(Y),
 Y,\bP)$. Moreover, it is decidable which of these cases holds and
 the
 corresponding operator is computable from $\psi$.
\end{enumerate}
\end{theor}
\section{Parity Games on Graphs and the Synthesis Problem}\label{sec:games}
In   subsection \ref{sec:paritygames}, we provide standard
definitions and facts about infinite two-player perfect
information  games on graphs. In \cite{BL69},  a reduction of the
Church synthesis problem to infinite two-player games  on  finite
graphs
 was provided.  In subsection \ref{sect:games-cp}, we provide  a reduction of the Church
   synthesis problem  with parameters to
 infinite two-player games  on infinite graphs;  this reduction is
 ``uniform" in the parameters.
The main  definability results needed for the proof of  Theorem
\ref{th:main-new} are given in Sect. \ref{sect:definable} and
Sect. \ref{sec:interpretation}.

\subsection{Parity Games on Graphs}\label{sec:paritygames}
We consider here two-player perfect information  games, played on
graphs, in which each player chooses, in turn, a vertex adjacent
to a current vertex. The presentation is based on \cite{PP04}.

A (directed) bipartite graph $G=(V_1,V_2, E)$ is called a
\emph{game arena} if the outdegree of every vertex is at least
one. If $G$ is an arena, a game on $G$ is defined by an initial
node $v_{init}\in V_1$ and a set of winning $\omega$-paths $\fair$
from this node.

Player  I plays on vertices in $V_1$ and Player II on vertices in
$V_2$. A play  from a node $v_1=v_{init}$ is an infinite  path
$v_1v_2  \dots v_i \dots $ in $G$  formed by the two players
starting from  the initial position  $v_1$. Whenever the current
position $v_i$ belongs to $V_1$ (respectively $V_2$), then Player
I (respectively, Player II)   chooses a successor node $v_{i+1} $
such  $E(v_i,v_{i+1})$. Since the graph is bipartite, Player I
plays at the odd positions ($v_{2i+1}\in V_i$) and Player II plays
at the even positions ($v_{2i}\in V_2 $).
 Player I wins if the
play belongs to $\fair$.

A strategy  $f$ for Player I (Player II) is a function  which
assigns to every  path of even (respectively, odd) length  a node
adjacent to the last node of the path. A play $v_{init} v_2v_3
\dots $ is played according to a strategy $f_1$ of Player I
(strategy $f_2$ of  Player II) if for every prefix $\pi=v_{init}
v_2 \dots v_n$ of even (respectively, odd) length
$v_{n+1}=f_1(\pi)$ (respectively, $v_{n+1}=f_2(\pi)$). A strategy
is winning for Player I (respectively, for Player II) if all the
plays played according to this strategy are in $\fair$
(respectively, in the complement of $\fair$). A strategy is
\emph{memoryless} if it depends only on the last nodes in the
path.

Parity games are games on graphs in which the set of winning paths
are defined by parity conditions. More precisely, let $G=(V_1,V_2,
E)$  be a game arena and let $c\:V_1\cup V_2\ra \{0,1,\dots m\}$
be a coloring.

Let $\rho=v_1v_2 \dots $ be a play.  With such a play $\rho$,  we
associate the set  of colors $C_{\rho}$ that appear infinitely
many times in the $\omega$-sequence $\col(v_1)col(v_2)\dots$;  a
play  $\rho$ is winning  for Player I  if the minimal element of
$C_{\rho} $ is odd. The following theorem due to Emerson/Julta and
Mostowski (see, \cite{EJ91,GTW02,PP04}) is fundamental:
\begin{theor} \label{th:par} In a parity game, one of the players has a
memoryless winning strategy.
\end{theor}

\subsection{Games and the Church Synthesis
Problem}\label{sect:games-cp} Let
${\aut}=\tuple{\autstates,\Sigma,\tra, \qinit, {\col}}$ be a
deterministic parity automaton over the alphabet
$\Sigma=\{0,1\}\times \{0,1\}\times \{0,1\}$, let  $R_{\aut}
\subseteq \{0,1\}^{\omega} \times \{0,1\}^{\omega} \times
\{0,1\}^{\omega} $ be the relation defined by $\aut$  and let
$\bP$ be a subset of $\nat$. We will define a parity game
$G_{\aut,\bP}$ such that
\begin{enumerate}[(1)]
\item  Player I has a winning strategy in  $G_{\aut,\bP}$  iff
there is a \sco operator $G\:\{0,1\}^{\omega}\ra
\{0,1\}^{\omega}$ such that $\neg R_{\aut}(G(Y),Y,\bP)$ holds for
every $Y$.
\item  Player II has a winning strategy in  $G_{\aut,\bP}$  iff
there is a  \retrospective operator $F\:\{0,1\}^{\omega}\ra
\{0,1\}^{\omega}$ such that $ R_{\aut}(X,F(X),\bP)$ holds for
every $X$.
\end{enumerate}
The arena $G(V_1,V_2,E)$ of $G_{\aut,\bP}$  is defined as follows:
\begin{enumerate}[{\ }]
\item{\bf Nodes:} $V_1={\autstates} \times \nat$ and $V_2=\autstates \times \{0,1\}
\times
\nat$.
\item{\bf Edges from $V_1$:}
 From  $\tuple{q,n}\in V_1$   two  edges exit; one to $\tuple{q,0,n}\in V_2$,
 and the second to  $\tuple{q,1,n}\in V_2$.
 We will assign  labels to these edges. The first one will be labeled
 by 0 and the second one will be labeled by 1.  These edge labels play no
 role in the game on our graph; however, it will be convenient   to refer to them later.
\item{\bf Edges from $V_2$:} From  $\tuple{q,a,n}\in
V_2$   two  edges  exit defined as follows: let $c$ be $1$ if
$n\in \bP$ and $0$ if $n\not\in\bP$; and for $b\in\{0,1\}$ let
$q_b$ be $\delta_{\aut}(q,\tuple{a,b,c})$. One edge from
$\tuple{q,a,n} $ is connected to $\tuple{q_0,n+1}$, and the second
one to $\tuple{q_1,n+1}$. We  label the first edge by 0, and the
second one by 1.
\end{enumerate}
The color of a node of  the arena is defined by the color of its
automaton's component, i.e.,
$c(\tuple{q,n})=c(\tuple{q,a,n}=\col(q)$.

The  node $\tuple{\qinit,0}$ is the initial node of the game.

Every node of the game graph for $G_{\aut,\bP}$ has two
successors. The subsets of $V_1$ (respectively, of $V_2$) can be
identified with the memoryless strategies of Player I
(respectively, of Player II). For a subset $U_1\subseteq V_1$, the
corresponding memoryless strategy $f_{U_1}$ is defined as
$$f_{U_1}(\tuple{q,n})=\left\{
\begin{array}{ll}\tuple{q,1,n} & \mbox{ if }\tuple{q,n}\in U_1\\
\tuple{q,0,n} & \mbox{otherwise}
\end{array}
\right.
$$
In other words, for $v\in V_1$ the strategy $f_{U_1}$ chooses the
nodes reachable from $v$ by the edge with  the label $U_1(v)$.
\begin{observation}[bijection between the memoryless strategies and the
subset of nodes]\label{lem:biject} The function that assigns to
every subset $U$ of $V_1$ the strategy $f_U$ for Player I is a
bijection between the set of memoryless strategies  for Player I
and the subset of $V_1$. Similarly, the function that assigns to
every subset $U$ of $V_2$ the strategy $f_U$ is a  bijection
between the set of memoryless strategies  for Player II  and the
subset of $V_2$.
\end{observation}
A subset $U_1\subseteq V_1$, induces a function
$h_{U_1}\:\{0,1\}^*\ra V_2$ and a \sco operator
$F_{U_1}\:\{0,1\}^{\omega}\ra \{0,1\}^{\omega}$. First, we
provide the definition for $h_{U_1}$, and  later for $F_{U_1}$.

 Let $G_{U_1}$ be
the subgraph of $G_{\aut,\bP}$, obtained by removing from every
node $v\in V_1$ the edge labelled by $\neg U_1(v)$, and removing
the label from the other edge exiting $v$. In this graph, every
$V_1$ node has outdegree one, and every $V_2$ node has two exiting
edges; one is labeled by 0 and the other is  labeled by 1. For
every $\pi$
 in $\{0,1\}^*$ there is a unique path from $\tuple{q_{init},0}$ to a
state $v_2\in V_2$ such that $\pi$ is the sequence  of labels on
the edges of this path; this node $v_2$ is $h_{U_1}$ image of
$\pi$.

Now a \sco operator $F_{U_1}\:\{0,1\}^{\omega}\ra
\{0,1\}^{\omega}$ induced by  $U_1$ is defined as follows. Let
$\pi =b_0b_1\dots $ be an $\omega$-string. There is a unique
$\omega$-path $\rho$ from $\tuple{q_{init},0}$ in $G_{U_1}$ such
that $\pi$ is the sequence labels on the edges of this path. Let
$v_1 v_2 \dots$ be the sequence of $V_1$ nodes on $\rho$ and let
$a_i=1$ if $v_i\in U_1$ and 0 otherwise. The $\omega$ sequence
$a_0a_1\dots$ is defined as the $F_{U_1} $ image of $\pi$.

Similarly, $U_2\subseteq V_2$  induces  a function
$h_{U_2}\:\{0,1\}^+\ra V_1$  and \retrospective operator
$F_{U_2}$.

Below we often use  ``a function $F$ corresponds to a set $U$" as
synonym  ``a set $U$ induces a function  $F$".

The properties of the above constructions are summarized as
follows:
\begin{lemma}\label{lem:un}
\begin{enumerate}[\em(1)]
\item Let $U_1$ be a subset of    $V_1$.
The memoryless strategy defined by $U_1$ is winning for Player I
in  $G_{\aut,\bP}$ iff $\neg R_{\aut}(F_{U_1}(Y),Y,\bP)$ holds for
every $Y$.
\item Let $U_2$ be a subset of $V_2$.
The memoryless strategy defined by $U_2$ is winning for Player II
in  $G_{\aut,\bP}$ iff $ R_{\aut}(X,F_{U_2}(X),\bP)$ holds for
every $X$.
\item Let $\vp(X,Y,Z)$ be an $\MLO$ formula equivalent to $\aut$, let $\psi$ be $\vp(X,Y,P) $ and let
$M$ be $\tuple{\nat,<,\bP}$. The  memoryless strategy defined by
$U$ is winning  for Player I (respectively, Player II) in
$G_{\aut,\bP}$ iff the operator induced by $U$ is a winning
strategy for Player I (respectively, for Player II) in
$\mG^M_\psi$.
\end{enumerate}
\end{lemma}
Our next objective is to show that the set of memoryless winning
strategies and the operator induced by a memoryless  strategy are
$\MLO$-definable in $\tuple{\nat,<,\bP}$. For this purpose, we
 show in \ref{sect:definable} that these are definable in the
monadic-second order logic for the structure appropriate for the
game graph $G_{\aut,\bP}$. Then, in Sect.
\ref{sec:interpretation}, we translate these definitions to $\MLO$
formulas over $\tuple{\nat,<,\bP}$.
 \subsection{Definability in the game structure}
\label{sect:definable}
The game arena $G_{\aut,\bP}$ can be
considered as a logical structure $M=M_{\aut,\bP}$  for the
signature $\tau_{\aut}=\{R_i\:i\in \autstates\cup
\autstates\times\{0,1\} \}\cup\{\mathit{Init},P,\prec,E_0,E_1\}$,
where $R_i$, $\mathit{Init}$ and $P$ are unary predicates and
$\prec,~,E_0,~E_1$ are binary predicates with the interpretation
$$R^M_i=\left\{\begin{array}{ll}
\{\tuple {q,j}\:j\in  \nat \}& \mbox{for }i= q\in \autstates\\
\{\tuple {q,a,j}\:j\in  \nat \}& \mbox{for }i= \tuple{q,a} \in
\autstates\times\{0,1\} \end{array} \right.
$$
$$P^M=\{\tuple{q,m}\:m\in \bP\}\cup \{\tuple{q,a,m}\:a\in \{0,1\} \mbox{ and}~m\in
\bP\}$$
$$\mathit{Init}^M=\{\tuple{\qinit,0}\}$$
$$\mathbf{E}^M_0(v_1,v_2) \mbox{~(respectively,
}\mathbf{E}^M_1(v_1,v_2))\mbox{ holds}
$$$$
 \mbox{ iff there is an edge labeled by 0 (respectively, by 1)
from $v_1$ to $v_2$}.$$

$$v_1\prec v_2 \mbox{ iff } v_1=\tuple{i_1,j_1} \mbox{ and }
v_2=\tuple{i_2,j_2} \mbox{ and } j_1<j_2.$$

The next lemma shows that the set of memoryless winning strategies
is definable in $M_{\aut,\bP}$.
\begin{lemma}[The set of memoryless winning strategies is definable in
$M_{\aut,\bP}$]\label{lem:winG}Let $\aut$ be a parity automaton.
\begin{enumerate}[\em(1)]
\item There is a monadic second-order formula $\Wing(X)$ such that
$M_{\aut,\bP}\models\Wing(U)$  iff  $ U $ corresponds to  a
memoryless winning strategy for  Player I.
\item There is a monadic second-order formula $\Loseg(X)$ such that
$M_{\aut,\bP}\models\Loseg(U)$  iff $ U $ corresponds to a
memoryless winning strategy for  Player II. \item Moreover,
$\Wing(X)$ and  $\Loseg(X)$ are computable from $\aut$.

\end{enumerate}
\end{lemma}
\proof We will formalize that  ``player I wins all the plays
consistent with  a memoryless strategy $X$".

A play is an infinite path   that starts
 from the initial node.
  Note that the arena $G_{\aut,\bP}$ is an acyclic graph.
  Hence, we can formalize that ``$Z$ is the set of nodes of an infinite path
  that starts from a node $v$" as a formula $\play(v,Z)$ which is
  the conjunction of the following formulas:
  \begin{enumerate}[(1)]
    \item For every node $u$ of $Z$ there is a unique node $u'\in Z$  such that an edge from $u$ enters
  $u'$.
  \item For every node $u\neq v$ of $Z$ there is a unique node $u'\in Z$  such that an edge from $u'$ enters $u$;
  there is no edge that enters from a node of $Z$ into $v$.
  \item $v$ is in $Z$.
  \item For every partition of $Z$ into two non-empty set $Z_1$ and
  $Z_2$ there is an edge between a node in $Z_1$ and a node in
  $Z_2$.
  \end{enumerate}
 The assertion  ``$u$ is a Player I node" is formalized by the formula
  $\mathit{Pos}_1(u)$, defined as $\bigvee_{q\in\autstates}R_q(u)$.
  Next, we formalize that ``$Z$ is the set of nodes of a play
  consistent with a strategy $U$ of Player I" by the formula
  $\mathit{Consis}(U,Z)$ which is the conjunction of the following
  formulas:
  \begin{enumerate}[(1)]
  \item ``$U$ is a subset of Player I nodes"
  is formalized by
  $$\forall u (u\in U\ra \mathit{Pos}_1(u))$$
\item ``$Z$ is the set of nodes of  a play from the initial node":
$$\exists v (Init(v)\wedge \play(v,Z))$$
\item ``$Z$ is consistent with $U$"
$$\forall zz' \in Z\big(\mathit{Pos}_1(z)\wedge(E_1(z,z')\vee E_1(z.z'))\big)\ra (E_1(z,z') \leftrightarrow z\in U)$$
    \end{enumerate}

Assume that the coloring function of $\aut$ assigns to the states
numbers  in the set $\{0, 1 \dots ,m\}$, and let $Q_i$ ($i=0,\dots
m$)  be the set of states of $\aut$ which are colored by  $i$.
Color $i$ appears infinitely often in a play with the set of nodes
$Z$ if $\mathit{inf}_i(Z)$ defined as $\forall z \exists z'\in Z
(z\prec z'\wedge (\vee_{q\in Q_i}R_q(z'))) $ holds. Hence, the
formula $\mathit{Even}(Z)$ defined as
$$\bigvee_{k\leq m/2}
\big(\mathit{inf}_{2k}(Z)\wedge\bigwedge_{j<2k}\neg
{inf}_{2k}(Z)\big)$$ holds for a play $Z$ iff the minimal color
that appears infinitely often in $Z$ is even.

Finally, $\Wing(X)$ can be defined as $ \forall Z
(\mathit{Consis}(X,Z)\ra  \neg \mathit{Even}(Z))$.

Note that our construction of $\Wing(X)$  from $\aut$ is
algorithmic.

The formula  $\Loseg(X)$ is defined from $\aut$ similarly. \qed
Next, we will show that the operator $F_{U}\:\{0,1\}^{\omega}\ra
\{0,1\}^{\omega}$ induced by a memoryless
  strategy $U$ of one of the players is definable  in $M_{\aut,\bP}$.
 The operator $F_{U}$ maps $\omega$-strings to $\omega$ strings.
 Therefore, we should agree how the $\omega$-strings are encoded by
 subset of $M_{\aut,\bP}$.

 Note that for $q\in \autstates$ the set of elements in $R_q$ ordered
 by $\prec$ is isomorphic to $\tuple{\nat,<}$. Accordingly, we can
 represent the $\omega$-strings by the subsets of $R_{qinit}$,
and
 in the next lemma the operators from $\{0,1\}^{\omega}$ to $ \{0,1\}^{\omega}$
 are identified with corresponding functions from the set of subset of $R_{\qinit}$ to the set of subset of $R_{\qinit}$.
 \begin{lemma}[Definability of the operator induced by a memoryless
 strategy] \label{lem:def-oper}
 Let $\aut$ be a parity automaton.
\begin{enumerate}[\em(1)]
\item There is a monadic second-order formula $\psi_I(X,Y,U)$ such that
$M_{\aut,\bP}\models \psi_I(\bX,\bY,\bU)$   iff  $ \bU $ is
 a memoryless  strategy for
Player I in $G_{\aut,\bP}$, $\bX,\bY\subseteq R_{\qinit}$   and
$\bX=F_\bU(\bY)$, where $F_\bU$ is the operator induced by $\bU$.
\item There is a monadic second-order formula $\psi_{II}(X,Y,U)$ such that
$M_{\aut,\bP}\models \psi_{II}(\bX,\bY,\bU)$   iff  $ \bU $ is
 a memoryless  strategy for
Player II in $G_{\aut,\bP}$, $\bX,\bY\subseteq R_{\qinit}$   and
$\bY=F_\bU(\bX)$, where $F_\bU$ is the operator induced by $\bU$.
\item
Moreover,  $\psi_I$ and $\psi_{II}$  are computable from $\aut$.\qed
\end{enumerate}
\end{lemma}
\proof We just formalize in the monadic-second order logic the
construction  of $F_U$ given in Sect. \ref{sect:games-cp}.

Let
  $\mathit{Consis}(U,Z)$ be the formula from the proof of Lemma
  \ref{lem:winG} which expresses
 ``$Z$ is the set of nodes of a play
  consistent with a strategy $U$ of Player I".
 We need to say that $X$ (respectively, $Y$) is the sequence of edges\footnote{Strictly speaking,
 the sequence of labels of the edges.}  chosen by Player I (respectively, by Player II)
  in the  play $Z$.
It can be formalized by  formula $\mathit{moves}(X,Y,Z)$ which is
the conjunction of
\begin{enumerate}[(1)]
\item $X\subseteq  R_{\qinit}\
\wedge Y\subseteq  R_{\qinit}$ - $X$ and $Y$ ``encodes" $\omega$
strings over $\{0,1\}$.
\item  We can formalize that $X$ is the sequence of edges chosen
by Player I  as follows. Let  $u\in Z$ be a  node of  Player II.
Then, $u$ is $\tuple{q,1,j}$ iff $\tuple{\qinit,j}$ is in $X$:
$$\forall u\in Z\forall v\in X \big(Pos_{II}(u)\wedge \neg(u\prec
v)\wedge\neg(v\prec u)\big)\ra \big(v\in X\leftrightarrow
\bigvee_{q\in \autstates} u\in R_{q,1}\big)$$
\item Similarly, we can formalize that $Y$ is the sequence of edges chosen
by Player II  as follows. Let $u=\tuple{q,a,j}\in Z$ and $
u'=\tuple{q',j+1}\in Z$ and let $c$ be 1 (respectively, $c$ be 0)
if $j\in \bP$ (respectively, $j\not\in\bP$) and let $b$ be 1
(respectively, $b$ be 0) if $\tuple{\qinit,j}\in Y$ (respectively,
$\tuple{\qinit,j}\not\in Y$). Then,
$q'=\delta_{\aut}(q,\tuple{a,b,c})$.
\end{enumerate}
Note that for each memoryless strategy $U$ of Player I, we have
$$\forall Y\subseteq  R_{\qinit} \exists !Z \exists! X
\mathit{Consis}(U,Z)\wedge \mathit{moves}(X,Y,Z) $$. Finally,
$\psi_I(X,Y,U)$  can be defined as $\exists Z
\mathit{Consis}(U,Z)\wedge \mathit{moves}(X,Y,Z) $.

$\psi_{II}(X,Y,U)$  is defined in a similar way.\qed

\subsection{Interpretation of $M_{\aut,\bP}$ in the structure
$\tuple{\nat,<\bP}$}\label{sec:interpretation}
 Every  set $S $ of nodes in $M_{\aut,\bP}$
corresponds to the tuple $\tuple{ \dots , W_i,\dots}$ where $i\in
\autstates\cup \autstates\times\{0,1\}$) of subsets of $\nat$,
such that $\tuple{q,m}\in S \mbox{ iff } m\in W_q$ and
$\tuple{q,a,m} \in S $ iff $ m\in W_{\tuple{q,a}}$.

The  proof of the following lemma shows that there is an
interpretation of the structure $M_{\aut,\bP}$ in the structure
$\tuple{\nat,<\bP}$.
\begin{lemma}\label{lem:trans} For every formula
$\psi(X^1,\dots X^k)$ in the second order monadic logic over the
signature $\tau_{\aut}$ with  free monadic variables $X^1,\dots,
X^k$ there is a  formula $\varphi(Y,\overrightarrow{Z^1}\dots
,\overrightarrow{Z^k})$, where $\overrightarrow{Z^j}$  is a tuple
of monadic variables $\{Z^j_i\:i\in \autstates\cup
\autstates\times\{0,1\}\}$, such that for every $\bP
\subseteq\nat$ and  a tuple $\tuple{ \dots ,W^j_i,\dots}$ where
$j\in\{1,\dots ,k\}$  and $i\in \autstates\cup
\autstates\times\{0,1\}$) of subsets of $\nat$ the following
equivalence holds:
 $$\tuple{\nat,<}\models \varphi(\bP, \dots ,W^j_i,\dots) ,
 \mbox{ iff }M_{\aut,\bP}\models\psi(S^1,\dots, S^k)$$
 $$\mbox{where $S^j$ is the   subset of nodes
 in }G_{\aut,\bP}, \mbox{ which corresponds to }\tuple{ \dots,
W^j_i,\dots}.$$ Moreover, there is an algorithms that computes
$\vphi$ from $\psi$.
\end{lemma}
\begin{proof} The proof proceeds by the structural induction.
It is more convenient to consider the first-order version of the
monadic second-order logic (see, Subsection \ref{sec:fov}).

{\em Basis} - Atomic formulas.
\begin{enumerate}[$\bullet$]
\item
$Sing(X)$ is translated as
$$\bigvee_{ i\in
\autstates\cup \autstates\times\{0,1\}}Sing(Z_i) \wedge
\bigwedge_{ i \neq i' \in \autstates\cup
\autstates\times\{0,1\}}(Sing(Z_i)\ra Empty(Z_{i'})),
$$
where $Empty(W)$ is a shorthand for $\forall W'(W\subseteq W')$.

\item $X^1\subseteq X^2$ is translated as
$$\bigwedge_{ i\in \autstates\cup \autstates\times\{0,1\}}
Z^1_i\subseteq Z^2_i$$
\item $X^1\prec X^2$ is translated as the conjunction of the
following:
\begin{enumerate}[(1)]
\item one of the $Z^1_i$  $( i\in \autstates\cup
\autstates\times\{0,1\}$) is singleton and all the others are
empty.
\item one of the $Z^2_i$  $( i\in \autstates\cup
\autstates\times\{0,1\}$) is singleton and all the others are
empty.
\item if $Z^1_{i_1}$  and $Z^2_{i_2}$ are singletons, then $Z^1_{i_1}<Z^2_{i_2}
$, i.e., the  (unique) natural number which is in $Z^1_{i_1}$ is
less than the natural number in $Z^2_{i_2}$.
\end{enumerate}

\item Other
relations in $\tau_{\aut}$ are translated similarly just by the
straightforward formalization of our definition of the game arena,
e.g., $E_0(X^1, X^2)$ is translated as the conjunction of the
following:
\begin{enumerate}[(1)]
\item one of the $Z^1_i$  $( i\in \autstates\cup
\autstates\times\{0,1\}$) is singleton and all others are empty.
\item one of the $Z^2_i$  $( i\in \autstates\cup
\autstates\times\{0,1\}$) is singleton and all others are empty.
\item if $Z^1_{q}$ (for some $q\in \autstates$)  is singleton,  then    $Z^2_{\tuple{q,0}}$   is singleton and both of them
contain the same natural number.

\item  if $Z^1_{\tuple{q,a}}$  (for some $q\in \autstates$ and $a\in \{0,1,\}$)
is singleton, and the unique number $n$ from  $Z^1_{\tuple{q,a}}$
is in $\bP$ (i.e., $Z^1_{\tuple{q,a}}\subseteq \bP$) and $q'$ is
be $\delta_{\aut}(q,\tuple{a,0,1})$, then $Z^2_{{q'}}$ is
singleton and the unique element of $Z^2_{{q'}}$ is the successor
of the unique element of $Z^1_{\tuple{q,a}}$.
\item  if $Z^1_{\tuple{q,a}}$  (for some $q\in \autstates$ and $a\in \{0,1,\}$)
is singleton, and the unique number $n$ from  $Z^1_{\tuple{q,a}}$
is not in $\bP$ (i.e., $\neg(Z^1_{\tuple{q,a}}\subseteq \bP)$) and
$q'$ is  $\delta_{\aut}(q,\tuple{a,0,0})$, then $Z^2_{{q'}}$ is
singleton, and the unique element of $Z^2_{{q'}}$ is the successor
of the  unique element of $Z^1_{\tuple{q,a}}$.
\end{enumerate}
\end{enumerate}
{\em Inductive step.} If $\vphi_j$ is the translation of $\psi_j$
for $j=1,2$, then $\psi_1\wedge \psi_2$ is translated as
$\vphi_1\wedge \vphi_2$ and $\neg \psi_1$ is translated as
$\neg\vphi_1$.

Finally, if  $\varphi(Y,\overrightarrow{Z^1}\dots
,\overrightarrow{Z^k})$ is the translation of $\psi(X^1,\dots
X^k)$, then $\exists X^1 \psi $ is translated as $\exists
\overrightarrow{Z^1}\vphi$.
\end{proof}
The next two lemmas use  the interpretation of $M_{\aut,\bP}$  in
$\tuple{Nat,<\bP}$ to show $\MLO$ definability in
$\tuple{Nat,<\bP}$ of   the set of memoryless winning strategies
and the operator induced by a memoryless strategy.
\begin{lemma}[The set of memoryless winning strategies is definable in
$\tuple{Nat,<\bP}$] \label{lem:winform} $~$\\
 Let ${\aut}=\tuple{\autstates,\Sigma,\tra, q_{init},
{\col}}$, be a deterministic parity automaton over the alphabet
$\Sigma=\{0,1\}\times \{0,1\}\times \{0,1\}$.
\begin{enumerate}[\em(1)]
\item There is an $\MLO$ formula $\win_{\aut}(Z_1, \dots,
Z_{|\autstates|}, Z)$ such that for every  $\bP \subseteq\nat$ and
 $W_1,\dots,
W_{|\autstates|}\subseteq \nat$:
$$\nat\models \win_{\aut}(W_1,\dots ,W_{|\autstates|}, \bP)$$
 iff the corresponding subset
$U\subseteq {\autstates} \times \nat$ defines a memoryless winning
strategy  for Player I in  $G_{\aut,\bP}$.
\item There is an $\MLO$ formula $\loose_{\aut}(Z_1, \dots,
Z_{|\autstates|}, Z'_1, \dots, Z'_{|\autstates|} ,Z)$ such that
for every  $\bP \subseteq\nat$ and $W_1,\dots,
W_{|\autstates|}, W'_1,\dots ,W'_{|\autstates|}\subseteq \nat$:
$$\nat\models \loose_{\aut}(W_1,
\dots, W_{|\autstates|},W'_1, \dots, W'_{|\autstates|} ,\bP)$$
iff the corresponding subset $U\subseteq {\autstates} \times
\{0,1\}\times \nat$ defines a  memoryless  winning  strategy
for Player II in $G_{\aut,\bP}$.
\item
Moreover, there is an algorithm that computes  formulas
$\win_{\aut}$ and $\loose_{\aut}$ from $\aut$.
\end{enumerate}
\end{lemma}
\proof
Follows from Lemma \ref{lem:winG} and Lemma \ref{lem:trans}. Let
$\Wing(X)$ be the formula constructed in  Lemma \ref{lem:winG}
which defines  (in $M_{\aut,\bP}$) the set of memoryless winning
strategies of the first player. Let $\psi(Y,\vec{Z^1})$ be its
translation, as in Lemma \ref{lem:trans}, where
$\overrightarrow{Z^1} $ is the tuple of variables indexed by $i\in
\autstates\cup \autstates\times\{0,1\}$. Note that since
$\overrightarrow{Z^1} $ defines (in $M_{\aut,\bP}$)  a strategy of
Player I, $Z_{\tuple{q,a}}$  for $q\in \autstates$ and
$a\in\{0,1\}$ should be interpreted as the empty set. The formula
$\win_{\aut}$ is obtained from $\psi$  by replacing
$Z_{\tuple{q,a}}$  for $q\in \autstates$ and $a\in\{0,1\}$  by the
empty set and replacing $Y$ by $Z$.

$\loose_{\aut}$ is defined in a similar way.\qed

From Lemma \ref{lem:def-oper} and Lemma \ref{lem:trans},  by the
same arguments, we can derive the following Lemma:

\begin{lemma}[The operator induced by a memoryless strategy is definable in
$\tuple{Nat,<\bP}$] \label{lem:oper-def-new} $~$\\ Let
${\aut}=\tuple{\autstates,\Sigma,\tra, q_{init}, {\col}}$, be a
deterministic parity automaton over the alphabet
$\Sigma=\{0,1\}\times \{0,1\}\times \{0,1\}$.
\begin{enumerate}[\em(1)]
\item There is an $\MLO$ formula $\opI(X,Y,Z_1, \dots,
Z_{|\autstates|}, Z)$  which has the following property:
\begin{quote}
Let $\bP \subseteq\nat$ and
 $W_1,\dots,
W_{|\autstates|}\subseteq \nat$,  and let $F_U$ be the C-operator,
induced by  Player I memoryless strategy $ U\subseteq {\autstates}
\times \nat$$ $ which corresponds to $\tuple{W_1,\dots,
W_{|\autstates|}}$. Then, $\nat\models\opI(X,Y,W_1,\dots,
W_{|\autstates|},\bP)$ iff $X=F_U(Y)$.
\end{quote}

\item There is an $\MLO$ formula $\opS{\aut}(X,Y,Z_1, \dots,
Z_{|\autstates|}, Z'_1, \dots, Z'_{|\autstates|} ,Z)$ which has
the following property:
\begin{quote}
Let   $\bP \subseteq\nat$ and
 $W_1,\dots,
W_{|\autstates|}, W'_1,\dots ,W'_{|\autstates|}\subseteq \nat$ and
 let $F_U$ be the C-operator, induced  by Player II memoryless
strategy $U\subseteq {\autstates} \times \{0,1\}\times \nat$ which
corresponds to $\tuple{W_1, \dots, W_{|\autstates|},W'_1, \dots,
W'_{|\autstates|}}$. Then, $$\nat\models \opS(X,Y,W_1, \dots,
W_{|\autstates|},W'_1, \dots, W'_{|\autstates|} ,\bP)\mbox{
 iff }Y=F_U(X).$$
 \end{quote}

\item
Moreover, there is an algorithm that computes  formulas $\opI$ and
$\opS$ from $\aut$.
\end{enumerate}
\end{lemma}
\section{Proof of Theorem \ref{th:main-new}}\label{sec:th-main-new}
We are now almost ready to prove Theorem \ref{th:main-new}.
 We  will show an  algorithm that, given a formula
$\vp(X,Y,P)$, constructs \begin{enumerate}[(1)] \item a sentence
$\mathit{\WIN_\vp}(P)$ and \item formulas
$\mathit{St^I_\vp}(X,Y,P)$ and $\mathit{St^{II}_\vp}(X,Y,P)$
\end{enumerate}
  such
that for every structure $M= \tuple{\modP}$, Player II has a
winning strategy in  the games $\mG_\vp^M$ iff $M\models
\mathit{\WIN_\vp}$. If Player II (respectively, Player I) has a
winning strategy, then
 $\mathit{St^{II}_\vp}(X,Y,P)$  (respectively,
 $\mathit{St^I_\vp}(X,Y,P)$)
defines his winning strategy.

Let $\vp(X,Y,P)$  be a formula. We are going to construct
$\mathit{\WIN_\vp}(P)$, as follows. First, let $\vp'(X,Y,Z)$ be a
formula obtained from $\vp$ by replacing all the  occurrences of
$P$ by a fresh variable $Z$.

Let ${\aut}\!=\!\tuple{\autstates,\Sigma,\tra, q_{init}, {\col}}$ be a
deterministic parity automaton over the alphabet $\Sigma\!=\!\{0,1\}\times
\{0,1\}\times \{0,1\}$, which is equivalent to $\vp'$. Let
$\win_{\aut}(Z_1, \dots, Z_{|\autstates|}, Z)$ be constructed from
$\aut$, as in Lemma \ref{lem:winform}. Player I has a winning strategy
iff
\[M\models \exists Z_1\dots \exists Z_{|\autstates|}
\win_{\aut}(Z_1, \dots, Z_{|\autstates|}, P)\ .
\]
  Finally, $\mathit{\WIN_\vp}(P)$ can be defined as $\neg\exists
Z_1\dots \exists Z_{|\autstates|} \win_{\aut}(Z_1, \dots,
Z_{|\autstates|}, P)$.

The correctness of the construction follows from Lemma
\ref{lem:winform} and Lemma \ref{lem:un}.

In order to construct $\mathit{St^I_\vp}(X,Y,P)$,
$\mathit{St^{II}_\vp}(X,Y,P)$, we need  the following definition
and Theorem.

\begin{definition}[Selection]\label{dfn:selection}
 Let $\vphi(\bar{Y})$, $\psi(\bar{Y})$ be formulas and $C$ a class of structures.
We say that $\psi$ \emph{selects} (or, is a \emph{selector} for)
$\vphi$ \emph{over} $C$ iff for every $M \in C$:
\begin{enumerate}[(1)]
\item $M \models \exists^{\le 1}\bar{Y}\psi(\bar{Y})$,
\item $M \models \forall \bar{Y}(\psi(\bar{Y})\rar \vphi(\bar{Y}))$, and
\item $M \models \exists \bar{Y}\vphi(\bar{Y}) \rar \exists \bar{Y}\psi(\bar{Y})$.
\end{enumerate}
\end{definition}
Here, $\bar{Y}$ is a tuple of distinct variables and
``$\exists^{\le 1}\bar{Y}\ldots$'' stands for ``there exists at
most one...''. The definition can be rephrased as $\psi$ is a
selector of $\vp$ over $C$ iff for  each $M\in C$, if  $\vp$ is
satisfiable in $M$, then it is satisfiable by the (unique) tuple
defined by $\psi$.


We say that $C$ has the \emph{selection property} iff every
formula $\vphi$ has a selector $\psi$ over $C$.

\begin{theorem}\label{th:selection}
The class of labelled $\omega$-chains has the selection property.
Moreover,  there is an algorithm which constructs for every
$\vphi$ a formula $\psi$ which  selects $\vphi$ over the class of
labelled $\omega$-chains.
\end{theorem}
Theorem \ref{th:selection} was proved in \cite{R07}. Its version
without  the ``Moreover" clause  was stated without proof in
\cite{LS98}.

Now, we are ready to define $\mathit{St^I_\vp}(X,Y,P)$ and
$\mathit{St^{II}_\vp}(X,Y,P)$.

 Let
${\aut}=\tuple{\autstates,\Sigma,\tra, q_{init}, {\col}}$ and
$\win_{\aut}(Z_1, \dots, Z_{|\autstates|}, Z)$ be as in the
construction of   $\mathit{\WIN_\vp}(P)$ above.

Construct  $\alpha(Z_1, \dots, Z_{|\autstates|}, P)$ as a selector
for $\win_{\aut}(Z_1, \dots, Z_{|\autstates|}, P )$ over the class
$\{\tuple{\nat,<,\bP}\: \bP\subseteq \nat\}$ of structures. If
 $\win_{\aut}(Z_1, \dots, Z_{|\autstates|},
P )$ is satisfiable in $M$, then every tuple which satisfies it
corresponds  to  a memoryless winning strategy of Player I in the
parity game $G_{\aut,\bP}$. In particular, the tuple defined by
$\alpha$ describes a memoryless winning strategy of Player I in
the parity game $G_{\aut,\bP}$.  Let $\opI$ be as in Lemma
\ref{lem:winform}. Then, $\exists Z_1 \dots\exists
Z_{|\autstates|}\big(\alpha(Z_1, \dots, Z_{|\autstates|}, P)\wedge
\opI( X,Y,Z_1, \dots, Z_{|\autstates|}\big)$ defines in $M$ a
SC-operator $F$, induced by a memoryless winning strategy of
Player I in the parity game $G_{\aut,\bP}$.

Hence, by Lemma \ref{lem:un}, $\neg R_{\aut}(F(Y),Y,\bP)$ holds
for every $Y$. Therefore, by the definitions of $\aut$ and
$R_{\aut}$, we have
$$\tuple{\nat,<}\models \forall Y\neg\psi((F(Y),Y,\bP).$$
Hence $F$, defined by $\exists Z_1 \dots\exists
Z_{|\autstates|}\big(\alpha\wedge \opI( X,Y,Z_1, \dots,
Z_{|\autstates|}\big)$, is a winning strategy for Player I in
$\mG_\vp^M$.

The formula  $\mathit{St^{II}_\vp}(X,Y,P)$ which defines a winning
strategy for Player II   is constructed similarly.

\section{Conclusion and Related Work}\label{sec:conc}
We investigated the Church synthesis problem with parameters. We
provided the necessary and sufficient conditions for the
computability of Synthesis problems 1-3.

The conditions of Theorem \ref{th:reduction} and Theorem
\ref{th:condprob5}
 are sufficient, but are not necessary for the
computability of Synthesis problems 4-5. For example, let  $\fAct
=\{n!\:n\in\nat\}$ be the set of factorial numbers.  We can show
that  Problems 4-5 are computable for this predicate  $\fAct$
\cite{Rab06}.

 It is an open question whether the decidability of
$\tuple{\nat,<,\bP}$ is a sufficient condition for the
computability of Synthesis  problems 4-5.

We proved that the definability and synthesis parts of the
B\"{u}chi and Landweber theorem hold for all  expansions  of
$\omega$ by unary predicates.

B\"{u}chi proved  that the $\MLO$-theory of any countable ordinal
is decidable.
 After stating their main theorem,
 B\"{u}chi and Landweber write:
\begin{quote}
 ``We hope to present elsewhere a corresponding extension of [our main theorem] from $\om$
to any countable ordinal.''
\end{quote}
 However,  despite the fundamental role of the Church problem, no such extension is even mentioned in
 a later book by
 B\"{u}chi and Siefkes
 \cite{buchi:siefkes}, which summarizes the  theory of finite automata  and $\MLO$ over words of countable
 length.

In \cite{Rab06a}, we proved that the determinacy and decidability
parts of the B\"{u}chi and Landweber  theorem hold for all
countable ordinals;  however, its  definability and synthesis
parts hold for an ordinal $\alpha$ iff $\alpha<\om^\om$
\cite{RS06}.

In \cite{RT07}, the Church Problem for fragments of $\MLO$ was
considered. First-order $\MLO$ formulas  are $\MLO$ formulas
without the second-order quantifiers. In \cite{RT06}, it was
proved that if $\psi(X,Y)$ is a first-order $\MLO$ formula, then
one of the players has a first-order definable winning strategy in
$\mG^\om_\psi$. Similar results were obtained for several
interesting fragments of $\MLO$. However, it is an open question
whether these results hold when parameters are added.

Kupferman and Vardi \cite{KV97} considered the synthesis problem
with incomplete information  for the  specifications described by
 temporal logics  LTL and CTL$^*$.  Their main results deal with  the complexity of this synthesis problem.
 The decidability of the
 synthesis problem with incomplete information for LTL (respectively, for CTL$^*$) can be
 easily derived from the  B\"{u}chi-Landweber  (respectively, Rabin) theorem.  It seems
 that there are no interesting connections between the synthesis
 problems with incomplete information and the synthesis problems
 with parameters considered here.

In \cite{RT98} a program for  the   relativization of finite
automata theory was proposed. Our results can be seen as the first
step in this direction. This step  corresponds to the case where
oracles are \retrospective operators without inputs.

In the rest of this section we comment on Rabin's proof of the
Church  synthesis Problem, discuss possibilities of extending it
to the Church  synthesis Problem with parameters,  and state some
open questions.

 Rabin
\cite{Rab72} provided an alternative proof for computability of
the Church synthesis problem. This proof uses an automata on
infinite trees as a natural tool for treating the synthesis
problem.  A \retrospective operator $F\:\{0,1\}^{\omega} \ra
\{0,1\}^{\omega} $  can be represented by a labelled infinite full
binary tree $\tuple{T_2,<,\bS}$, where $\bS$ is a subset of the
tree nodes. Namely, the branches of the tree represent $X\in
\{0,1\}^{\omega} $ and the sequence of values assigned by $\bS$ to
the nodes along the branch $X$ represents $F(X)=Y\in
\{0,1\}^{\omega} $. Also, the fact that $\bS$ represents a
\retrospective  operator $F$ which uniformizes $\varphi(X,Y)$ can
be expressed  by an $\MLO$ formula $\psi(Z)$
 (computable from $\varphi(X,Y)$): $T_2\models \psi(\bS)$
 iff  $\nat  \models \forall \varphi(X,F_{\bS} (X))$ , where $F_{\bS}$ is the \retrospective operator that
  corresponds to $\bS$. Hence, the question whether there exists a \retrospective operator which uniformizes
 $\varphi$
  is reduced to the problem whether $T_2\models \exists Z
  \psi(Z)$. Now, the Rabin basis theorem states that if $T_2\models \exists Z
  \psi(Z)$ then there is a regular subset $\bS\subseteq T_2$ such
  that $T_2\models \psi(\bS)$. The \retrospective operator which
  corresponds to a regular set $\bS$ is computable by a finite
  state automaton. Hence, the  B\"{u}chi  and Landweber theorem is
  obtained as a consequence of the decidability of the monadic
  logic of order of  the full binary tree and the basis theorem.

  One could try to apply the Rabin method to the Church synthesis
  problem with parameters.
  The reduction which is similar to  Rabin's reduction shows
  that for every $\vp(X,Y,P)$ there is a sentence  $\mathit{WIN^{II}_\vp}(Q)$
  such that for every structure $M= \tuple{\modP}$ Player II wins
the games $\mG_\vp^M$ iff $$\tuple{T_2,<,\bQ} \models
\mathit{WIN^{II}_\vp}(Q),$$ where a node is in $\bQ$ if its
distance from
  the root is in $\bP$. Moreover, $\mathit{WIN^{II}_\vp}(Q)$ is
  computable from $\vp$ (cf. Theorem \ref{th:main-new}).

   Hence,  the decidability  of Problem 1$'$    for $\bP\subseteq \nat$  (recall that this is Problem 1 without its  constructive part, see Sect.
   \ref{sec:th2}) is reduced to the
  decidability of the monadic theory of the labelled full binary tree
  $\tuple{T_2,<,\bQ}$, where a node is in $\bQ$, if its distance from
  the root is in $\bP$. The decidability of the latter problem
  can be reduced by  Shelah-Stupp Muchnick Theorem \cite{She75,Wal02,Th03} to the decidability of $\tuple{\nat,<,\bP}$.
   Now, in order to establish  computability of problems 1-3, one can
   try to
   prove    the  basis theorem for $\tuple{T_2,<,\bQ}$.
   Unfortunately,
arguments similar to the proof of Theorem \ref{th:reduction} show
that  for $\bP,\bQ$, as above,  the following are equivalent:
\begin{enumerate}[(1)]
\item $\bP$ is ultimately periodic.
\item
 For
every $\psi(Z,Q)$ such that $T_2\models\exists Z\psi(Z,\bQ)$ there
is a finite state operator $F(Y,U)$ such that the  set which
corresponds to the \retrospective operator $\lambda X F(X,\bP)$
satisfies $\psi(Z,\bQ)$.
\end{enumerate}

%

In case selector over $\tuple{T_2,<,\bQ}$ is computable, it would
be easy to derive the computability of Problems 1-3.
 However, even the following is open

\noindent{\bf Open Question - Selectoion property for   an
expansion of $T_2$ by unary predicate.} \begin{quote} Is it true
that for every $\bQ\subseteq T_2$ every formula $\vp(Y,Q)$ has a
selector over $\tuple{T_2,<,\bQ}$, i.e., there is $\psi(Y,Q)$ such
that (1) $\tuple{T_2,<.\bQ} \models \exists^{\le 1}{Y}\psi({Y})$,
(2) $\tuple{T_2,<,\bQ} \models \forall {Y}(\psi({Y})\rar
\vphi({Y}))$, and (3) $\tuple{T_2,<.\bQ} \models \exists
{Y}\vphi({Y}) \rar \exists \bar{Y}\psi({Y})$.
\end{quote}

Note that this question  asks whether every expansion of $T_2$ has
the selection property and   is different from the Rabin's
uniformization problem  over $T_2$, which asks whether every
formula has a selector over the class of all the expansions of
$T_2$ by unary predicates. The negative answer to the Rabin
uniformization problem was obtained by Gurevich and Shelah
\cite{GS} who proved that  a formula $\vp(Y,Q)$ which express that
``if $Q$ is not empty than $Y$ is a singleton set which is a
subset of $Q$" has no selector over the class of all the
expansions of $T_2$.

We believe that  if the answer to the above mentioned question  is
positive, then its proof is non-trivial.

Our proof of Theorem \ref{th:intr1} implies that for $\bP\subseteq
\nat $ and $\bQ\subseteq T_2$,  where a node is in $\bQ$ if its
distance from
  the root is in $\bP$, the
following are equivalent:
\begin{enumerate}[(1)]
\item The  monadic theory of $\tuple{\nat,<,\bP}$ is
decidable,
\item
 For
every $\psi(Z,Q)$ such that $T_2\models\exists Z\psi(Z,\bQ)$ there
is a recursive  $\bS\subset T_2$ such that
$T_2\models\psi(\bS,\bQ)$.

\end{enumerate}

However, we do not know the answer to the following question:

\noindent{\bf Open Question:} Are the following assertions
equivalent?
\begin{enumerate}[(1)]
\item The monadic theory of $\tuple{T_2,<\bS}$ is decidable.
\item For
every $\psi(Z,U)$ such that $T_2\models\exists Z\psi(Z,\bS)$ there
is a  recursive set $\bQ\subset T_2$ such that
$T_2\models\psi(\bQ,\bS)$.
\end{enumerate}
%

 \section*{Acknowledgments} %
 I am grateful to the anonymous referees for their suggestions.


\begin{thebibliography}{Maz89a}
\bibitem[Bu60]{Bu}
J.~R. B\"{u}chi.
\newblock On a decision method in restricted second order arithmetic
\newblock In {\em Proc. International Congress on Logic, Methodology
and Philosophy of Science}, E. Nagel at al. eds, Stanford
University Press, pp 1-11, 1960.

\bibitem[BL69]{BL69} J. R.  B{\"u}chi and
               L. H. Landweber.
 Solving sequential conditions by finitestate strategies. Transactions of the AMS, 138(27):295--311, 1969.

 \bibitem[BS73]{buchi:siefkes} J. R. B\"{u}chi, D. Siefkes, The Monadic Second-order Theory of all Countable Ordinals, Springer Lecture Notes 328 (1973),
pp. 1-126.


\bibitem[CT02]{CT02}
  {O. Carton and
               W.Thomas}.
  The Monadic Theory of Morphic Infinite Words and
  Generalizations.
  {Inf. Comput.
    176}(1), pp. {51-65},
     {2002}.
     \bibitem[Ch69]{Ch69} Y. Choueka. Finite Automata on Infinite Structure.
Ph.D Thesis, Hebrew University, 1970.
     \bibitem[ER66]{ER66}{ C. Elgot and
               M. O. Rabin}.
 {Decidability and Undecidability of Extensions of Second
               (First) Order Theory of (Generalized) Successor.}
 {J. Symb. Log.},
  {31}(2),
  pp. {169-181},
{1966}.
\bibitem[EJ91]{EJ91}
E. A. Emerson, C. S. Jutla: Tree Automata, Mu-Calculus and
Determinacy (Extended Abstract) FOCS91: 368-377, 1991.
\bibitem[GTW02]{GTW02}
 E. Gr{\"a}del,
               W. Thomas and
               T. Wilke.
  {Automata, Logics, and Infinite Games},
  LNCS 2500,
 {2002}.

\bibitem[GS83]{GS}
  Y. Gurevich and S. Shelah.
  \newblock Rabin's uniformization problem.
  \newblock The Journal of Symbolic Logic, \textbf{48}:1105-1119, 1983.


\bibitem[KV97]{KV97} O. Kupferman and M.Y. Vardi,
        Synthesis with incomplete information,
   In 2nd International Conference on Temporal Logic,
       pp 91--106,
        1997.
\bibitem[Mc66]{mcnaughton} R. McNaughton, Testing and generating infinite sequences by a finite automaton, Information
and Control 9 (1966), pp. 521-530.
\bibitem[LS98]{LS98} S. Lifsches, S. Shelah, Uniformization and skolem functions in the class of trees,
Jou. of Symolic Logic, Vol. 63(1) (Mar. 1998), pp. 103-127.


\bibitem[PP04]{PP04} D.  Perrin and J. E. Pin.
Infinite Words Automata, Semigroups, Logic and Games.
 Pure and Applied Mathematics Vol 141
Elsevier, 2004.



\bibitem[Rab72]{Rab72}  M.~O. Rabin.  Automata on Infinite Objects and Church's Problem
 Amer. Math. Soc. Providence, RI, 1972.



\bibitem[Rab05]{R05}
 A. Rabinovich.  On decidability of monadic logic of order over the naturals extended by monadic predicates.
 2005 Summer Meeting of the Association for
Symbolic Logic, Logic Colloquium '05. The Bulletin of Symbolic
Logic {\bf 12}:343-344, 2006.


\bibitem[Rab06]{Rab06}
A. Rabinovich. The Church problem over $\omega$ expanded by
factorial numbers.  In preparation, 2006.

\bibitem[Rab06a]{Rab06a}
A. Rabinovich. The Church Problem  for Countable Ordinals.
Submitted, 2006.
\bibitem[Rab07]{R07}
A. Rabinovich.
 On decidability of Monadic logic of order over the
naturals extended by monadic predicates.
 Information and Computation, 2007.
\bibitem[RT98]{RT98}        A. Rabinovich and B.A. Trakhtenbrot.
From Finite Automata toward Hybrid Systems Proceddings of
Fundamentals of Computation Theory. Lecture Notes in Computer
Science 1450, pp. 411-422, Springer, 1998.
\bibitem[RS06]{RS06} A. Rabinovich and A. Shomrat. Selection in the Monadic Theory of Countable
Ordinals. Submitted 2006.
\bibitem[RT06]{RT06} A. Rabinovich and  W.~Thomas.
Decidable Theories of the Ordering of Natural Numbers with Unary
Predicates. In CSL 2006, Springer LNCS 4207, 562-574, 2006.


\bibitem[RT07]{RT07} A. Rabinovich and  W.~Thomas. Logical
Refinements of Church's Problem. In CSL 2007, LNCS 4646, 69-83,
2007.
\bibitem[Rob58]{Rob58} R. M.
Robinson.  Restricted Set-Theoretical Definitions in Arithmetic.
In Proceedings of the AMS Vol. 9, No. 2.  pp. 238-242, 1958.
\bibitem[Sem84]{Sem} A. Semenov. Logical theories of one-place functions
on the set of natural numbers. Mathematics of the USSR - Izvestia,
vol. 22, pp 587-618, 1984.

\bibitem[Se04]{Se04}
O. Serre. Games With Winning Conditions of High Borel Complexity.
In  ICALP 2004, LNCS volume 3142, pp. 1150-1162, 2004.


\bibitem[Shel75]{She75}
S.~Shelah. The monadic theory of order. \emph{Ann. of
Math.}~\textbf{102}:379--419, 1975.
\bibitem[Sie75]{Sie75} D. Siefkes. The recursive sets in certain monadic second order fragments of arithmetic.
 Arch. Math. Logik, pp71-80, 17(1975).

\bibitem[Th75]{Th75}
W.~Thomas. Das Entscheidungsproblem f\"{u}r einige Erweiterungen
der Nachfalger-Arithmetic. Ph. D. Thesis Albert-Ludwigs
Universit\"{a}t, 1975.
\bibitem[Th95]{Th95} W. Thomas. On the synthesis of
strategies in infinite games. In  STACS '95, LNCS vo. 900, pp.
1-13.  1995.
\bibitem[Th03]{Th03} W. Thomas. Constructing infinite graphs
with a decidable MSO-theory. In  MFCS03, LNCS 2747, 2003.
\bibitem[Trak61]{Tra61}
B.~A.~Trakhtenbrot.
\newblock Finite automata and the logic of one-place predicates. (Russian
version 1961). \newblock In AMS Transl. 59, 1966, pp. 23-55.
\bibitem[Wal02]{Wal02} I. Walukiewicz. Monadic second order logic on tree-like
structures.TCS 1:275, pp 311-346, 2002.
\end{thebibliography}
\end{document}